\documentclass{jfm}
\usepackage{graphicx}
\usepackage{epstopdf, epsfig}
\usepackage{amsmath,mathrsfs}
\usepackage[mathscr]{euscript}
\usepackage[usenames,dvipsnames]{xcolor}
\usepackage{appendix}
\usepackage[squaren,Gray,cdot]{SIunits}

\usepackage{comment}
\newcommand{\xhat}{\hat{x}}
\newcommand{\zhat}{\hat{z}}
\newcommand{\hhat}{\hat{h}}
\newcommand{\uhat}{\hat{u}}
\newcommand{\vhat}{\hat{v}}
\newcommand{\qhat}{\hat{q}}
\newcommand{\hPi}{\hat{\Pi}}
\newcommand{\htau}{\hat{\tau}}
\newcommand{\rmd}{\mathrm{d}}

\newcommand{\rmb}{\mathrm{b}}
\newcommand{\hzstar}{\hat{z}^{\ast}}

\shorttitle{Dip-coating with two liquids}
\shortauthor{L. Champougny et al.}
\title{Dip-coating flow in the presence of two immiscible liquids}
\author{Lor\`ene Champougny\aff{1},
  Benoit Scheid\aff{2},
  Alexander A. Korobkin\aff{3}
 \and Javier Rodr\'{\i}guez-Rodr\'{\i}guez\aff{1}
  \corresp{\email{bubbles@ing.uc3m.es}}}
\affiliation{\aff{1}Fluid Mechanics Group, Universidad Carlos III de Madrid, Legan\'es, Spain.
\aff{2}Transfers, Interfaces and Processes (TIPs), Universit\'e\, Libre de Bruxelles, Brussels, Belgium.
\aff{3}School of Mathematics, University of East Anglia, Norwich, UK.
}
\begin{document}
\maketitle
%
%
\begin{abstract}
Dip-coating is a common technique used to cover a solid surface with a thin liquid film, the thickness of which was successfully predicted by the theory developed by Landau \& Levich and Derjaguin in the 1940's.
In this work, we present an extension of their theory to the case where the dipping bath contains two immiscible liquids, one lighter than the other, resulting in the entrainment of two thin films on the substrate.
We report how the thicknesses of the coated films depend on the capillary number, on the ratios of the properties of the two liquids and on the relative thickness of the upper fluid layer in the bath.
We also show that the liquid/liquid and liquid/gas interfaces evolve independently from each other as if only one liquid was coated, except for a very small region where their separation falls quickly to its asymptotic value and the shear stresses at the two interfaces peak. 
Interestingly, we find that the final coated thicknesses are determined by the values of these maximum shear stresses.
\end{abstract}
%
\section{Introduction}
%
The process of dip-coating aims to deposit a thin liquid layer on the surface of an object by withdrawing the latter from a liquid bath in which it is initially immersed.
This simple technique is employed in many industrial processes in order to modify the properties of a solid surface \citep{Scriven1988}, with applications ranging from anti-corrosion treatments and optically modified glasses, to surface functionalisation of bio-implants.
The hydrodynamics of dip-coating have been known for nearly 80 years. 
The original description of this process dates back to the now classical work of \citet{LandauLevich1942}, followed by the contribution of \citet{Derjaguin1943}, who included the effect of gravity.
We will refer to their description of the one-liquid dip-coating problem as the LLD theory.
Later on, when the relevant mathematical tools became available, the rigorous asymptotic theory underlying these initial developments was proposed by \citet{WilsonJEM1982}.

Nurtured by practical applications, experimental and theoretical aspects of dip-coating have kept attracting scientific attention ever since, as highlighted by the recent review by \citet{RioBoulogne2017}.
New effects and regimes have been found to arise from the use of partially-wetting \citep{Snoeijer2008, Tewes2019} or textured \citep{Seiwert2011} solid substrates, non-Newtonian liquids \citep{Hurez1990, Maillard2016}, surface active molecules \citep{Park1991, Scheid2010, Champougny2015} or jammed hydrophobic micro-particles \citep{Dixit2013, Ouriemi2013} adsorbed at the liquid/gas interface.
The two latter cases can be regarded as stepping stones to multi-phase dip-coating situations.
Indeed, interfacial particle rafts were shown to behave as two-dimensional elastic sheets floating on top of the liquid bath \citep{Vella2004}, while surfactant monolayers at liquid/gas interfaces are known to exhibit analogs to two-dimensional liquid or gaseous phases \citep{Vollhardt2010}.

Beyond these analogies, the dip-coating configuration in which the bath contains two immiscible liquids, one floating on top of the other, has never been explored.
The objective of the present paper is to extend the LLD dip-coating theory to describe liquid entrainment at such a gas/liquid/liquid \emph{compound interface}.
Processes occurring at compound interfaces, for example made of a layer of oil floating on water, raise interest in the context of environmental science or semiconductor electronics.
The surface of the oceans can be seen as a compound interface, due to the presence of the sea surface microlayer \citep{Hardy1982, Liss2005} or even more so in the event of an oil spill \citep{Fingas2015}, hence the relevance of processes such as bubble bursting \citep{Feng2014, Stewart2015} or bouncing \citep{Feng2016} at gas/liquid/liquid interfaces. These are examples of elementary processes that occur at the millimetric or sub-millimetric scale in the ocean and whose physics is related to the one we consider in this work.
In a different context, dip-coating through a compound interface, consisting in lifting a substrate through a layer of carbon-nanotube-laden ink floating on top of a water bath, was experimentally investigated by \citet{Jinkins2017}.
They showed the potential of this method, known as floating evaporative self-assembly in the context of semiconductor electronics, for the deposition of well-aligned carbon nanotubes arrays.

In the present study, we will restrict ourselves to the situation in which both liquid phases are dragged, giving birth to a superposition of two liquid films on the substrate.
It is worth mentioning the connection of this configuration to the model introduced by \citet{Seiwert2011} to describe the dip-coating of a textured solid with a single liquid phase.
In that work, the effect of the texture was modelled as a secondary, uniform layer made of a fluid with a viscosity higher than that of the coating liquid.
Finally, in a different geometry, the entrainment of a thin liquid film at a liquid/liquid interface has been described in the context of lubricant-infused surfaces \citep{Li2019}.
When a water droplet is sliding on an oil-imbibed surface, thin oil films observed behind, beneath or around the droplet have been shown to obey similar scaling laws as in the LLD theory \citep{Daniel2017, Kreder2018}.

Our theoretical investigation of dip-coating through a gas/liquid/liquid compound interface, leading to the deposition of a double liquid coating, is organised as follows.
In section \ref{sec:model_description}, we develop the problem formulation and comment on the various dimensionless control parameters.
Our results are presented and discussed in sections \ref{sec:results1} and \ref{sec:results2}.
In section \ref{sec:results1}, we focus on the typical interfacial shapes and flow structures obtained from the numerical solutions of the model. 
These observations allow us to rationalise quantitatively the asymptotic film thicknesses coated on the substrate, highlighting the universality of the entrainment mechanism (viscous stresses \textit{vs} capillary suction).
In section \ref{sec:results2}, we examine the dependence of the film thicknesses on a number of dimensionless control parameters of the problem.
Importantly, we show that the existence of a physically-meaningful second film is limited to specific, finite regions in the parameter space.
In section \ref{sec:discussion}, we comment on the scope and limitations of our model in relation to disjoining pressure effects and partial wetting conditions between the two liquids.
Finally, section \ref{sec:conclusion} is devoted to the conclusions.
%
\section{Description of the model} \label{sec:model_description}
%
\subsection{Dimensional flow equations} \label{ssec:dimensional_formulation}
\begin{figure}
\centering
\includegraphics[width=8cm]{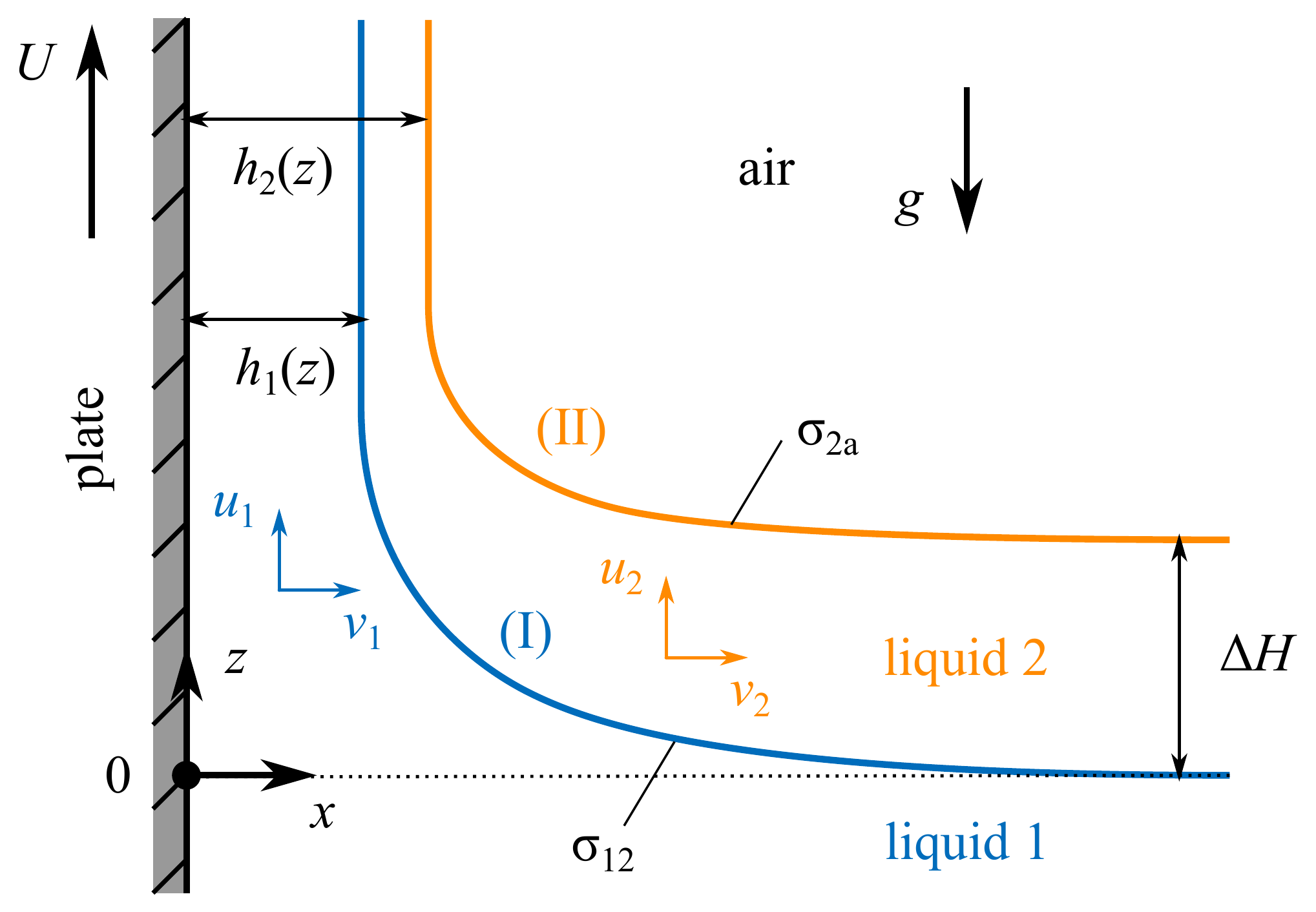}
\caption{Sketch of the flow configuration described in this work: a solid plate is pulled vertically at constant speed through a compound bath made of a lighter liquid (2) on top of a denser one (1). The liquid 1 / liquid 2 and liquid 2 / air interfaces are denoted by (I) and (II), respectively.}
\label{fig:sketch_coordinates_definitions}
\end{figure}
We present here the dimensional formulation of the two-liquid dip-coating problem sketched in figure~\ref{fig:sketch_coordinates_definitions}. 
A solid plate is vertically lifted at a constant speed $U$ through a stratified bath made of two immiscible liquid layers.
This bath consists of a pool of liquid 1 (with density $\rho_1$ and viscosity $\mu_1$), covered by a layer of liquid 2 (with density $\rho_2 < \rho_1$ and viscosity $\mu_2$), which has a thickness $\Delta H$ far from the plate.
The interface between liquid 1 and 2, denoted by (I), has an interfacial tension $\sigma_{12}$, while the interface between liquid 2 and the surrounding air, denoted by (II), has a surface tension $\sigma_{2a}$.
Both liquids 1 and 2 are supposed to perfectly wet the plate. However, at this stage, we do not impose perfect wetting conditions between liquids 1 and 2. 
In other words, the spreading factor $S = \sigma_{1a} - (\sigma_{2a} + \sigma_{12})$, where $\sigma_{1a}$ denotes the surface tension of liquid 1, can be either positive or negative for the moment (see \textit{e.g.} \citet{DeGennes2013}). 
We will discuss the effect of the wetting properties of liquid 2 on liquid 1 in § \ref{ssec:validity_conditions} and later on in § \ref{ssec:partial_wetting}.

We consider the problem to be \emph{steady} and two-dimensional in space, described by the horizontal coordinate $x$ and the vertical coordinate $z$ (see figure~\ref{fig:sketch_coordinates_definitions}).
As the plate crosses the compound bath, we assume it entrains two thin films: a lower film of liquid 1 with thickness $h_1(z)$ and an upper film of liquid 2 with thickness $\delta h (z) = h_2(z) - h_1(z)$. 
The velocity fields are $u_1 \boldsymbol{e_z} + v_1 \boldsymbol{e_x}$ and $u_2 \boldsymbol{e_z} + v_2 \boldsymbol{e_x}$ in liquids 1 and 2, respectively.

Following the approach by \citet{LandauLevich1942} and \citet{Derjaguin1943}, we describe the flow of each phase in the transition region, referred to as the \emph{dynamic meniscus}, that connects the films of constant thickness downstream to the \emph{static meniscus} upstream.
For each liquid, the vertical extent of the static meniscus is assumed much larger than the corresponding film thickness, allowing us to apply the lubrication theory to the flow.
In this steady lubrication approach, the momentum conservation equations in the $z$-direction read:
\begin{eqnarray}
0 & = & -\frac{\partial p_1}{\partial z} + \mu_1 \frac{\partial^2 u_{1}}{\partial x^2} - \rho_1 g,%
\qquad \qquad (0 < x < h_1(z), \quad z > 0) \label{eq:momentum_dim_1}\\
0 & = & -\frac{\partial p_2}{\partial z} +  \mu_2 \frac{\partial^2 u_{2}}{\partial x^2} - \rho_2 g.%
\qquad \qquad (h_1(z) < x < h_2(z), \quad z > \Delta H)\label{eq:momentum_dim_2}
\end{eqnarray}
The $x$-independent pressures $p_i$ ($i = 1, 2$) are related to the atmospheric pressure $p_a$ and to the interfacial curvatures $\kappa_i$ through
\begin{eqnarray}
p_1 &=& p_2 - \sigma_{12} \kappa_1, \qquad \qquad (0 < x < h_1(z), \quad z > 0) \label{eq:capillary_pressure_1}\\
p_2 &=& p_a - \sigma_{2a} \kappa_2, \qquad \qquad (h_1(z) < x < h_2(z), \quad z > \Delta H), \label{eq:capillary_pressure_2}\\
p_2 &=& p_a + \rho_2 g (\Delta H - z), \qquad \qquad (x > h_1(z), \quad 0 < z \le \Delta H),
\end{eqnarray}
where, for $i = 1, 2$,
\begin{equation} \label{eq:curvature_dim}
   \kappa_i = \frac{\partial^2h_i}{\partial z^2}\left[1 + \left(\frac{\partial h_i}{\partial z}\right)^2\right]^{-3/2}.
\end{equation}
Denoting $Q_1$ the flow rate of liquid 1 (\textit{i.e.} under interface (I)) and $Q_2$ the \emph{total} flow rate of liquids 1 and 2 (\textit{i.e.} under interface (II)), the quasi-steady thickness-averaged continuity equations read
\begin{equation}
\frac{\partial Q_i}{\partial z} = 0 \quad \quad \mathrm{for} \ i = 1, 2.
\label{eq:continuity_dim_1and2}
\end{equation}
The expressions relating the flow rates $Q_i$ with the flow velocities, $u_i$ and the locations of the interfaces $h_i$ will be provided below, once the dimensionless formulation is introduced.
%
\subsection{Non-dimensionalization} \label{ssec:non_dimensionalization}
%
A convenient choice for the length scale to non-dimensionalise the problem is the capillary length $\ell_c$ based on the properties of liquid 1 in the absence of the buoyancy effect created by liquid 2, defined as
\begin{equation}
    \ell_c = \sqrt{\frac{\sigma_{12}}{\rho_1 g}}.
    \label{eq:def_capillary_length}
\end{equation}
Note that $\ell_c$ neither represents the scale of the static meniscus of liquid 1, nor the one of liquid 2.
The ``physical'' dimensional capillary lengths associated to the menisci of liquids 1 and 2 are respectively $\ell_{c,1} = \sqrt{\sigma_{12}/(\rho_1-\rho_2)g}$ and $\ell_{c,2} = \sqrt{\sigma_{2a}/\rho_2 g}$, whose dimensionless versions indeed appear in the static configuration presented in § \ref{ssec:matching} (equations~\eqref{eq:kappa_stat_1} and \eqref{eq:kappa_stat_2}, respectively).
The independent and dependent variables of the problem are made dimensionless as follows:
\begin{eqnarray}
(x, z) & \quad\longrightarrow\quad & \ell_c \, (x, z) \\
(h_1,h_2) & \quad\longrightarrow\quad & \ell_c \, (h_1, h_2), \\
(p_1,p_2) & \quad\longrightarrow\quad & \sigma_{12}/\ell_c \, (p_1, p_2), \\
(u_1,u_2,v_1, v_2) & \quad\longrightarrow\quad & U \, (u_1, u_2,v_1, v_2).
\end{eqnarray}
The thickness of the layer of liquid 2 is also scaled by $\ell_c$, still keeping the same notation: $\Delta H \longrightarrow \ell_c \, \Delta H$. The capillary number is introduced as $Ca = \mu_1 U / \sigma_{12}$. Also, the following dimensionless parameters are defined:
\begin{equation}
    M = \frac{\mu_2}{\mu_1}, \qquad %
    \Sigma = \frac{\sigma_{2a}}{\sigma_{12}}, \qquad %
    R = \frac{\rho_2}{\rho_1}.
\end{equation}
Based on these definitions, the dimensionless momentum equations deduced from equations \eqref{eq:momentum_dim_1} and \eqref{eq:momentum_dim_2} read
\begin{eqnarray}
0 & = & \Pi_1 + Ca \, \frac{\partial^2 u_{1}}{\partial x^2}, %
\qquad \qquad (0 < x < h_1(z), \quad z > 0) \label{eq:momentum_adim_1} \\
0 & = & \Pi_2  + M \, Ca \, \frac{\partial^2 u_{2}}{\partial x^2}, %
\qquad \qquad (h_1(z) < x < h_2(z), \quad z > \Delta H) \label{eq:momentum_adim_2}
\end{eqnarray}
where the dimensionless pressure gradients are
\begin{eqnarray}
\Pi_1 & = & \frac{\partial \kappa_1}{\partial z} + \Sigma \, \frac{\partial \kappa_2}{\partial z} - 1, \qquad \qquad (0 < x < h_1(z), \quad z > \Delta H)
\label{eq:gradP_adim_1} \\
\Pi_2 & = & \Sigma \, \frac{\partial \kappa_2}{\partial z} - R , \qquad \qquad (h_1(z) < x < h_2(z), \quad  z > \Delta H). \label{eq:gradP_adim_2}
\end{eqnarray}
For $0 < z \le \Delta H$, equation (\ref{eq:gradP_adim_1}) must be replaced by
\begin{equation}
    \Pi_1 = \frac{\partial \kappa_1}{\partial z} - 1 + R.
\label{eq:gradP_adim_1_no_kappa_2}
\end{equation}
Notice that the pressure gradient $\Pi_1$ and its derivative are continuous at $z = \Delta H$, as $\Pi_2$ smoothly approaches 0 as $z \rightarrow 0^+$.
In the above expressions, $\kappa_1$ and $\kappa_2$ are the dimensionless curvatures, which are still given by \eqref{eq:curvature_dim} after non-dimensionalisation.
Equations \eqref{eq:momentum_adim_1} and \eqref{eq:momentum_adim_2} can be integrated with respect to $x$ across the corresponding layers, using the following boundary conditions: 
\begin{eqnarray}
\text{Stress-free at } x &=& h_2: \qquad \frac{\partial u_{2}}{\partial x} = 0, \label{eq:BCx_adim_1}\\
\text{Velocity continuity at } x &=& h_1: \qquad u_1 = u_2, \label{eq:BCx_adim_2}\\
\text{Stress continuity at } x &=& h_1: \qquad \frac{\partial u_{1}}{\partial x} = M \frac{\partial u_{2}}{\partial x}, \label{eq:BCx_adim_3}\\
\text{No slip at } x &=& 0:~~ \qquad u_1 = 1. \label{eq:BCx_adim_4}
\end{eqnarray}
After obtaining the vertical velocity fields $u_{1}$ and $u_{2}$ (whose expressions are given in appendix \ref{apdx:velocity_and_shear}), we compute the dimensionless flow rate $Q_1$ in the lower layer and the \emph{total} dimensionless flow rate $Q_2$ carried by the two layers:
\begin{equation} \label{eq:Q12}
Q_i = Ca \, \int_0^{h_i} u_i (x) \, \mathrm{d}x = Ca\,h_i + \Pi_1 \, F_{i1} + \Pi_2 \, F_{i2},
\end{equation}
with $i = \left\{1, 2\right\}$ and
\begin{eqnarray}
F_{11} & = & \frac{h_1^3}{3} \label{eq:F11}\\
F_{12} & = & \left(h_2-h_1\right)\frac{h_1^2}{2} \label{eq:F12}\\
F_{21} & = & \frac{h_1^3}{3} + \left(h_2-h_1\right)\frac{h_1^2}{2} \label{eq:F21}\\
F_{22} & = & \frac{\left(h_2-h_1\right)^3}{3M} + h_1\left(h_2-h_1\right)^2 + \left(h_2-h_1\right) \frac{h_1^2}{2}. \label{eq:F22}
\end{eqnarray}
Replacing the flow rates $Q_i$ given by \eqref{eq:Q12} in the continuity equations \eqref{eq:continuity_dim_1and2} -- which remain unchanged upon non-dimensionalization -- we arrive at a system of two fourth-order ordinary differential equations.
This system is closed by implementing eight boundary conditions as follows.
Far up from the reservoir, we impose that the thicknesses of the coated films converge asymptotically to constants: for $i = \left\{1, 2\right\}$, $\partial h_i/\partial z = \partial^2 h_i/\partial z^2 = 0$ when $z \rightarrow +\infty$. 
Towards the reservoir, the static menisci connect to the flat surfaces of liquids 1 and 2.
We therefore impose that $\kappa_1=0$ and $\partial h_1/\partial z \rightarrow \infty$ at $z = 0$ for liquid 1; and $\kappa_2 = 0$ and $\partial h_2/\partial z \rightarrow \infty$ at $z = \Delta H$ for liquid 2.
%
\subsection{Scaling and simplification} \label{ssec:rescaling}
%
The problem consisting of equations \eqref{eq:continuity_dim_1and2}, along with equations \eqref{eq:Q12} and the above-mentioned boundary conditions, could be solved numerically for $h_1$ and $h_2$.
However, in order to gain physical insight into the scalings governing the different regions of the flow, we further simplify the problem using the matched asymptotic expansion treatment proposed by \citet{WilsonJEM1982} in the case of the one-liquid LLD flow.

The film thicknesses $h_i$ and the vertical coordinate $z$ are rescaled with a dimensionless parameter $\varepsilon$ as $h_i = \varepsilon \hat{h}_i$ and $z = \varepsilon^{\alpha} \hat{z}$. This technique will allow us to find the spatial scales at which viscous stress and capillary pressure gradient are equally important or, in other words, to find the scale of the dynamic menisci. The small parameter $\varepsilon$ can be interpreted as the ratio between the length scale (in the direction perpendicular to the plate) at which viscous effect are important, and the capillary length.
Requiring that the curvatures $\kappa_i$ are of order unity (so dimensional curvatures are of order $\ell_c^{-1}$) when the films approach the static menisci, we find $\alpha=1/2$ \citep{WilsonJEM1982}. 
Expanding equations \eqref{eq:continuity_dim_1and2} and \eqref{eq:Q12} -- \eqref{eq:F22} for $\varepsilon \ll 1$ and retaining terms up to order $\varepsilon^{1/2}$ yields
\begin{equation}
\frac{\partial}{\partial \hat{z}} \left[\frac{Ca}{\varepsilon^{3/2}} \, \hat{h}_i\right. +%
\left(\frac{\partial^3\hat{h}_1}{\partial\hat{z}^3} +%
\Sigma \, \frac{\partial^3\hat{h}_2}{\partial\hat{z}^3}-\varepsilon^{1/2}\right)\hat{F}_{i1} +%
\left.\left(\Sigma \, \frac{\partial^3\hat{h}_2}{\partial\hat{z}^3} -%
\varepsilon^{1/2}R\right)\hat{F}_{i2}\right] = 0,%
\label{eq:thinfilm_with_delta} 
\end{equation}
\noindent where $\hat{F}_{ij}$ are the functions defined in equations \eqref{eq:F11} -- \eqref{eq:F22} but in terms of $\hat{h}_i$ instead of $h_i$. 
One assumption that we make here is that the dynamic menisci region is located above $z = \Delta H$. This means that the two interfaces $h_1(z)$ and $h_2(z)$ are defined there. Note that the curvatures are reduced to the second derivatives of the thicknesses, as the next term in their expansion is of order $\varepsilon$.

Analogously to the formulation developed in \citet{WilsonJEM1982} for the one-liquid case, equation \eqref{eq:thinfilm_with_delta} encompasses the effects of both capillary- and gravity-driven drainage on the steady state film thicknesses.
In this work, we will focus on the case where capillary effects prevail over gravity.
In equation \eqref{eq:thinfilm_with_delta}, viscous stresses will be of the same order as capillary ones if $\varepsilon = Ca^{2/3}$, which is the same scaling obtained in the one-liquid case by \citet{LandauLevich1942}.
Finally, at leading order, the equations satisfied by the thicknesses are
\begin{equation}
\frac{\partial}{\partial \hat{z}}
\left[\hat{h}_i + \frac{\partial^3 \hat{h}_1}{\partial \hat{z}^3}\hat{F}_{i1} +%
\Sigma \, \frac{\partial^3 \hat{h}_2}{\partial \hat{z}^3}\left(\hat{F}_{i1} + \hat{F}_{i2}\right)\right] = 0.\label{eq:thinfilm_eqns}
\end{equation}
The boundary conditions far from the liquid bath remain unchanged, as compared to the full formulation (developed in the last paragraph of \ref{ssec:non_dimensionalization}): $\hat{h}_i^{\prime} = \hat{h}_i^{\prime \prime} = 0$ for $\hat{z} \rightarrow + \infty$, where the prime denotes the derivative with respect to $\hat{z}$.
On the contrary, the boundary conditions near the reservoir have to be imposed through an asymptotic matching with the static menisci, as our ability to rigorously implement these boundary conditions at the liquid bath was lost when the curvature was linearized and gravity was neglected.
%
\subsection{Asymptotic matching with the static menisci} \label{ssec:matching}
%
\begin{figure}
\centering
\includegraphics[width=7cm]{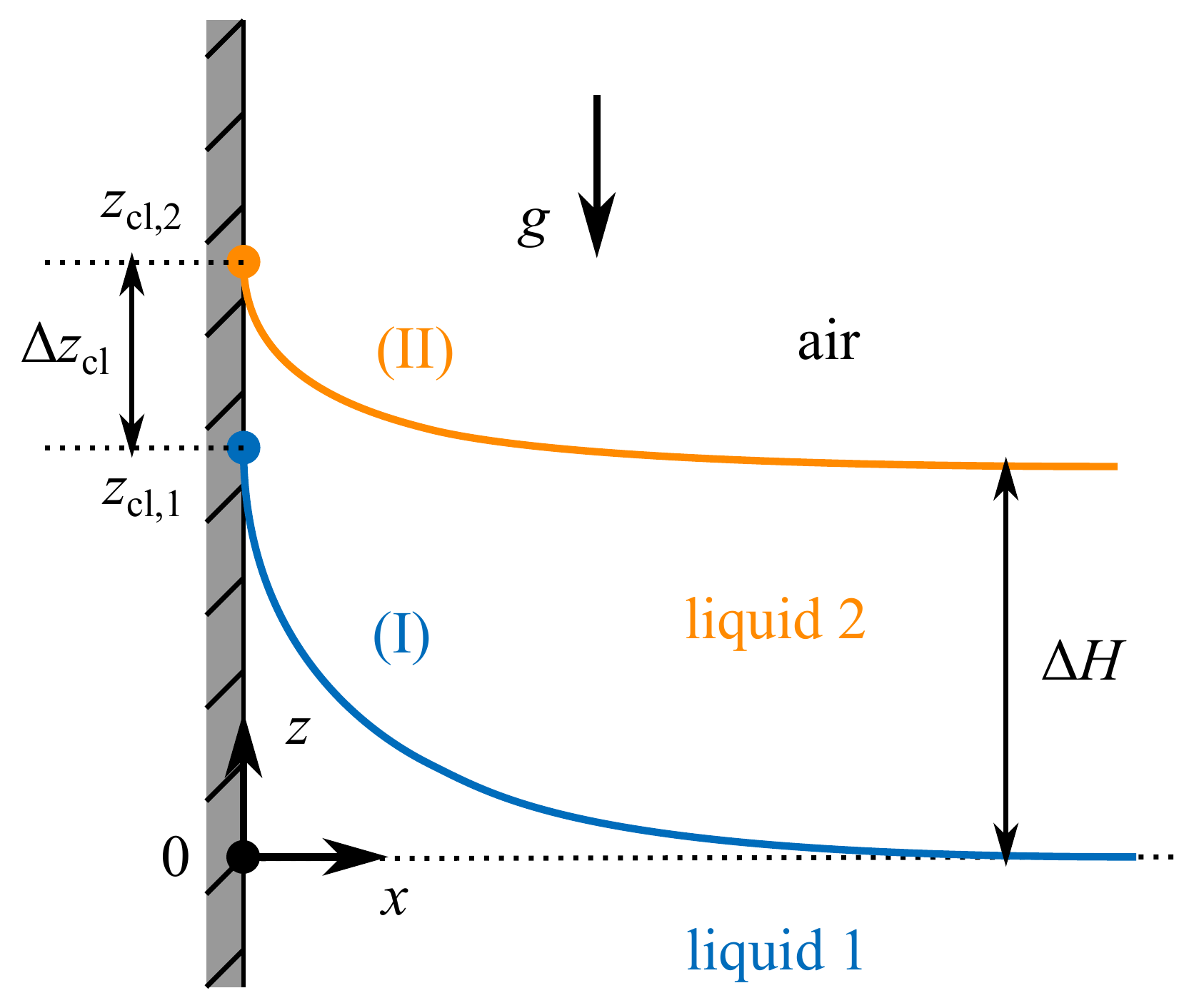}
\caption{Sketch of the static configuration: an immobile vertical plate is wetted by a compound bath at rest, made of a lighter liquid (2) on top of a denser one (1). The liquid 1 / liquid 2 and liquid 2 / air interfaces, denoted by (I) and (II), climb up to heights $z_{cl,1}$ and $z_{cl,2}$, respectively. The distance between the two corresponding contact lines on the plate is therefore $\Delta z_{cl} = z_{cl,2} - z_{cl,1}$.}
\label{fig:sketch_notations_static}
\end{figure}
%
%
\subsubsection*{Static configuration}
%
For the purpose of asymptotic matching, we briefly study the \emph{static} configuration depicted in figure~\ref{fig:sketch_notations_static}.
Setting $Ca = 0$ in equations \eqref{eq:momentum_adim_1} -- \eqref{eq:momentum_adim_2} (which also implies $Q_i = 0$) leads to $\Pi_1 = \Pi_2 = 0$, or equivalently:
\begin{align}
    \frac{\partial\kappa_1}{\partial z} + R - 1 = 0, \label{eq:meniscus_1} \\ 
    \Sigma \, \frac{\partial\kappa_2}{\partial z} - R = 0 \label{eq:meniscus_2}.
\end{align}
As done in the one-liquid case (see for instance \cite{LandauLifshitzBOOK}, Chapter 7, page 243), these equations can be integrated to find the exact solutions for the shape of the static menisci $h_1(z)$ and $h_2(z)$.
Since our purpose is to use these solutions to obtain asymptotic matching conditions for the dynamic thin film equations \eqref{eq:thinfilm_eqns}, we only need the approximated shape of the static interfaces (I) and (II) near the contact lines on the plate, \textit{i.e.} close to positions $z_{cl,1}$ and $z_{cl,2}$, respectively.

Because $h_1 (z_{cl,1}) = h_2 (z_{cl,2}) = 0$ (by definition) and  $h^{\prime}_1 (z_{cl,1}) = h^{\prime}_2 (z_{cl,2}) = 0$ (perfect wetting conditions on the plate), the Taylor expansions of the menisci profiles in the vicinity of the contact lines read, for $z \le z_{cl,2}$:
\begin{align}
h_1 & = \frac{1}{2}\kappa_1(z_{cl,1})\left(z - z_{cl,1}\right)^2, \label{eq:h1_parabola}\\
h_2 & = \frac{1}{2}\kappa_2(z_{cl,2})\left(z - z_{cl,2}\right)^2.\label{eq:h2_parabola}
\end{align}
By solving \eqref{eq:meniscus_1} and \eqref{eq:meniscus_2}, using the definition \eqref{eq:curvature_dim}, we obtain the expressions for the curvatures at the contact lines,
\begin{align}
    \kappa_1(z_{cl,1}) &= \sqrt{2(1-R)}  \quad \text{at} \quad z_{cl,1} = \sqrt{2/(1-R)}, \label{eq:kappa_stat_1} \\ 
    \kappa_2(z_{cl,2}) &= \sqrt{2R/\Sigma} \quad \text{at} \quad z_{cl,2} = \Delta H + \sqrt{2\Sigma/R}, \label{eq:kappa_stat_2}
\end{align}
for interfaces (I) and (II) respectively.
%
\subsubsection*{Asymptotic matching}
%
In the one-liquid case, thanks to the invariance of the problem in the $\hat{z}$-direction, imposing the value of the curvature in the limit $\hat{z} \rightarrow -\infty$ is enough to perform the asymptotic matching \citep{LandauLevich1942}.
In our two-liquid case, however, we benefit from the invariance along $\hat{z}$ only for one of the interfaces because the relative position of interfaces (I) and (II) far from the plate is fixed (they are separated by a given distance $\Delta H$).
Consequently, the boundary conditions in the limit $\hat{z} \rightarrow -\infty$ not only result from imposing the curvatures of the two static menisci, but also require specifying how far apart the interfaces are.
In order to fulfill these two conditions, the film thicknesses $\hat{h}_i$ are required to follow the parabolic approximations of the static menisci (Equations \eqref{eq:h1_parabola} and \eqref{eq:h2_parabola}), expressed in terms of the rescaled variables $\hat{h}_i = Ca^{-2/3} \, h_i$ and $\hat{z} = Ca^{-1/3} \, z$:
\begin{align}
\hat{h}_1 & = \frac{1}{2}\sqrt{2\left(1-R\right)} \left(\hat{z} - \hat{z}_{cl,1}\right)^2, \label{eq:eta1_parabola}\\
\hat{h}_2 & = \frac{1}{2}\sqrt{2R/\Sigma}\left(\hat{z} - \hat{z}_{cl,2}\right)^2,\label{eq:eta2_parabola}
\end{align}
in the limit $\hat{z} \rightarrow -\infty$. 
We have introduced $\hat{z}_{cl,i} = z_{cl,i} \, Ca^{-1/3}$, where $z_{cl,i}$ denotes the dimensionless location where interfaces (I) and (II) (for $i=1$ and $2$ respectively) meet the plate in the static configuration and for perfect wetting conditions.
Note that the matching between the dynamic and the static menisci will occur within a vertical distance of order $ \ell_c \, Ca^{1/3}$ from the contact lines of the hydrostatic solutions.
For $\varepsilon^{1/2} = Ca^{1/3} \ll 1$ (condition to neglect gravity in the dynamic menisci, see § \ref{ssec:rescaling} and § \ref{ssec:validity_conditions}), this distance is much smaller than the capillary length, therefore justifying that the static menisci can be approximated by parabolas.
%
\subsection{Dimensionless control parameters} \label{ssec:control_params}
%
\begin{table}
    \centering
    \begin{tabular}{cccc}
    Parameter & Definition & Value or range \\
    \hline
        Capillary number & $Ca = \mu_1 U / \sigma_{12}$ & $10^{-3}$ (fixed)\\
        floating layer thickness & $\quad \Delta H = \Delta H^{\rm (dim)} / \ell_c \quad$ & $2.57 - 5.20$\\
        Density ratio & $R = \rho_2/\rho_1$ & $0.885$ (fixed)\\
        Surface tension ratio & $\Sigma = \sigma_{2a}/\sigma_{12}$ & $0.333; \, 0.667; \,  1.0; \, 1.27$ \\
        Viscosity ratio & $M = \mu_2/\mu_1$ & $10^{-2} - 10^1$ \\
        \hline
    \end{tabular}
    \caption{Main dimensionless control parameters of the problem and corresponding values or ranges explored in this work.}
    \label{tab:param_table}
\end{table}
The problem formulated above depends on a large number of dimensionless control parameters: $Ca$, $R$, $\Sigma$, $M$, and $\Delta H$.
For the sake of conciseness, we will restrict our analysis to the parameters that are expected to have the largest impact on the flow structure and whose effect cannot be easily accounted for by some scaling argument.

We can estimate practically relevant orders of magnitude by considering a system made of a $40 \%$ glycerol aqueous solution (liquid 1) with a floating layer of silicone oil (liquid 2). 
The corresponding densities are $\rho_1 = 1100~\kilo\gram\per\meter^3$ for the glycerol solution \citep{Takamura2012gly} and $\rho_2 = 970~\kilo\gram\per\meter^3$ for the oil, yielding a density ratio $R = 0.885$.
The liquid 1/ liquid 2 interfacial tension and liquid 2/air surface tension are $\sigma_{12} = 30~\milli\newton\per\meter$ and $\sigma_{2a} = 20~\milli\newton\per\meter$, respectively, yielding $\Sigma = 0.667$.
Finally, the viscosity of the glycerol solution is $\mu_1 = 3.6~\milli\pascal\cdot\second$ \citep{Takamura2012gly} and silicone oils can have a viscosity $\mu_2$ ranging from a fraction of $\milli\pascal\cdot\second$ to hundreds of $\pascal\cdot\second$, while keeping their other properties (density and surface tension) roughly constant.
The values and ranges chosen in our computations for the dimensionless control parameters are summarized in Table~\ref{tab:param_table} and discussed in the following.
%
\subsubsection*{Capillary number $Ca$ and floating layer thickness $\Delta H$}
%
Importantly, after rescaling the problem as described in § \ref{ssec:rescaling}, the capillary number $Ca$ disappears from the formulation (see equations \eqref{eq:thinfilm_eqns}), except in the matching conditions \eqref{eq:eta1_parabola} -- \eqref{eq:eta2_parabola}.
In the static menisci, with which the matching is performed in the limit $\hat{z} \rightarrow -\infty$, the dimensionless thickness $\Delta H$ of the floating layer remains related to the rescaled distance $\Delta\hat{z}_{cl}$ between the contact lines of liquids 1 and 2 (see § \ref{ssec:matching}) through 
\begin{equation}
    \Delta\hat{z}_{cl}\,Ca^{1/3} = \Delta H + \sqrt{\frac{2\Sigma}{R}} - \sqrt{\frac{2}{1-R}}.
\label{eq:relationship_DeltaH_Ca}
\end{equation}
Thus, for given fluid properties, changing the capillary number $Ca$ has the same effect as changing the thickness $\Delta H$. 
For this reason, throughout this work, we will vary only $\Delta H$ while keeping $Ca = 10^{-3}$ constant.
This specific value was chosen in order to meet the condition $Ca^{1/3} \ll 1$, ensuring that gravity effects are negligible in the dynamic meniscus region, while being relevant in many applications \citep{RioBoulogne2017}.
The values of $\Delta H$ explored in this work correspond to the range where solutions with two coated films were found to exist, \textit{i.e.} between $\Delta H = 2.57$ and $5.20$ for our values of $Ca$, $\Sigma$ and $R$.
%
\subsubsection*{Density ratio $R$}
%
The density ratio will be set to $R = 0.885$ (corresponding to the glycerol solution/silicone oil system described above) throughout this study. 
We choose to keep this parameter constant since, in practice, $R$ will most of the time be of this order with aqueous solution/oil systems.
%
\subsubsection*{Surface tension ratio $\Sigma$}
%
Four values of the surface tension ratio will be explored: $\Sigma = 0.333$, $0.667$ (corresponding to the glycerol solution/silicone oil system described above), $1.0$ and $1.27$. 
These values are chosen around $1$, which is what can be achieved with most common fluids.
%
\subsubsection*{Viscosity ratio $M$}
%
For glycerol solution / silicone oil systems, the viscosities of both phases can be tuned over several orders of magnitude by varying the glycerol concentration in the water phase and changing the average chain length in the oil phase. 
Consequently, the values of viscosity ratio explored will span several decades, from $M = 10^{-2}$ to about $10^1$, the exact upper value being limited by the existence of solutions with two coated films, as will be shown in § \ref{ssec:thickness_maps}.
%
\subsection{Validity conditions of the model} \label{ssec:validity_conditions}
%
\subsubsection*{Stability of the hydrostatic configuration}
For the static configuration of figure~\ref{fig:sketch_notations_static} to exist in practice, we need the stability of the floating layer to be warranted. So far, no assumption has been made on the (dimensional) spreading coefficient $S = \sigma_{1a} - (\sigma_{2a} + \sigma_{12})$ of liquid 2 on liquid 1. In the case where $S>0$, the hydrostatic configuration where liquid 2 forms a continuous layer in the bath will be stable regardless of the thickness of this layer, at least as long as long range molecular forces (\textit{e.g.} van der Waals) can be neglected \citep{Leger1992}. This is the case for the $40~\%$ glycerol aqueous solution / silicone oil system described above, for which $\sigma_{1a} = 70~\milli\newton\per\meter$ \citep{Takamura2012gly}, leading to $S = 20~\milli\newton\per\meter > 0$.

In the case where $S<0$ however, the floating layer will be stable only above a certain critical thickness $\Delta H_c$ \citep{DeGennes2013}. This critical thickness, made dimensionless with the capillary length $\ell_c$, can be written with our notations
\begin{equation}
    \Delta H_c = \sqrt{\frac{-2\mathscr{S}}{R(1-R)}},
    \label{eq:floating_layer_DeltaHc}
\end{equation}
where $\mathscr{S} = S/\sigma_{12} < 0$ is the dimensionless spreading coefficient. For $\Delta H < \Delta H_c$, the floating layer is metastable and, if perturbed, will dewet \citep{Brochard1993}.
%
\subsubsection*{Negligible gravity}
We included gravity in our first, most general formulation developed in § \ref{ssec:dimensional_formulation} and § \ref{ssec:non_dimensionalization}. When performing the appropriate scaling in the dynamic menisci, retaining only leading order terms (see § \ref{ssec:rescaling}), we showed that the gravity term can be neglected provided $\varepsilon^{1/2} \ll 1$, that is to say $Ca^{1/3} \ll 1$. Note that this condition, which is met for our value of $Ca = 10^{-3}$, is the same as in the one-liquid LLD problem \citep{DeRyck1998}.
%
\subsubsection*{Negligible inertia}
Even at low capillary numbers, inertial effects may also arise when low-viscosity liquids are used. In our model, scaled as presented in § \ref{ssec:rescaling}, the conditions for inertia to be negligible as compared to viscous and capillary terms read
\begin{align}
     \text{in liquid 1} \qquad We &= Re \, Ca \ll 1, \label{eq:condition_Weber_1}\\
     \text{in liquid 2} \qquad We &= Re \, Ca \ll \mathrm{min}(\Sigma / R ; M/R), \label{eq:condition_Weber_2}
\end{align}
where $We$ is the Weber number \citep{RioBoulogne2017}, $Re = \rho_1 U \ell_c/ \mu_1$ the Reynolds number (based on the properties of liquid 1) and $Ca$ the capillary number (as defined in § \ref{ssec:non_dimensionalization}). Taking the above-mentioned $40 \%$ glycerol aqueous solution as liquid 1 and $Ca = 10^{-3}$, we find $We \approx 4 \times 10^{-3}$, showing that conditions \eqref{eq:condition_Weber_1} -- \eqref{eq:condition_Weber_2} are fulfilled in the whole range of parameters explored in our work (see table \ref{tab:param_table}).

Additionally, in the general formulation where gravity is still present, neglecting inertial forces as compared to gravitational ones in both liquid films requires $Re \, Ca^{2/3} \ll 1$, that is to say $We \ll  Ca^{1/3}$, which is fulfilled for our parameters. Note that the softer condition $Re \, Ca^{2/3} \sim 1$ (\textit{i.e.} $We \sim Ca^{1/3}$) is sufficient in the dynamic menisci, as gravity effects are already small there.
%
\section{Results: flow morphology and entrainment mechanism}
\label{sec:results1}
%
In this section, we will present and discuss the typical shape of interfaces (I) and (II) (§ \ref{ssec:shape_interfaces}), as well as the flow structure in the dynamic menisci (§ \ref{ssec:flow_structure}), obtained by solving the model described in section \ref{sec:model_description} using the pseudo-transient numerical strategy described in Appendix \ref{apdx:transient_and_numerics}.
Based on the observation of these results, we will propose a simplified, geometrical description (§ \ref{ssec:geometrical_model}) allowing us to derive scaling laws for the asymptotic thickness of the coated liquid films (§ \ref{ssec:scaling_laws}).
%
\subsection{Shape of the interfaces} \label{ssec:shape_interfaces}
%
\begin{figure}
\centering
\includegraphics[trim=2.5cm 1.5cm 2.5cm 1.5cm,clip,width=\linewidth]{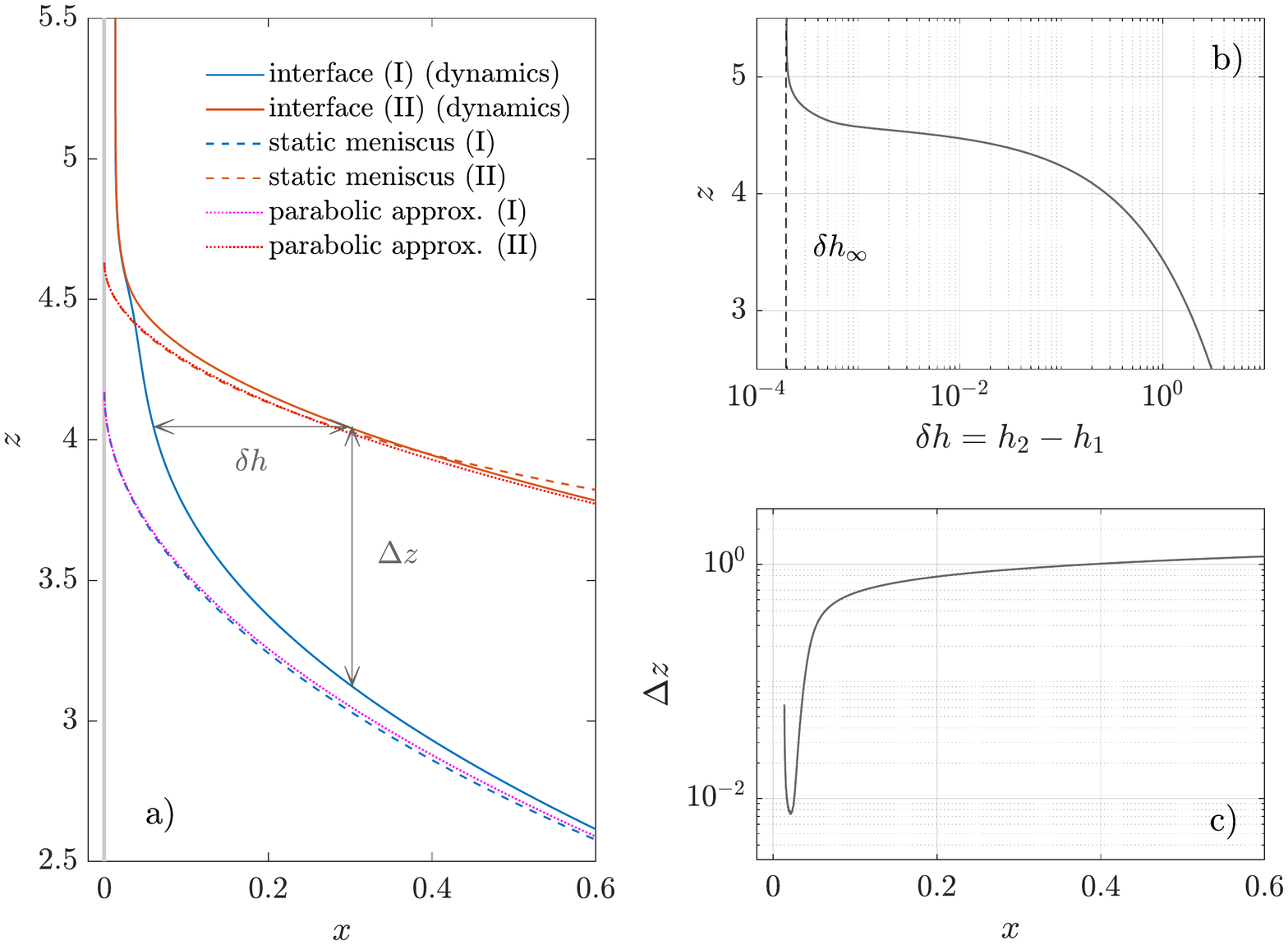}
\caption{Shape of the interfaces and matching to the static menisci for $\Sigma = 0.667$, $R = 0.885$, $M = 1$, $Ca = 10^{-3}$ and $\Delta H = 3.403$. --- a) In the dynamic menisci region, the interfaces (solid lines) depart from the static solutions (dashed lines) to connect to two thin films of uniform thicknesses (that can barely be distinguished at this scale). The dotted lines show the parabolic approximations of the static menisci near the plate, used in the matching conditions \eqref{eq:eta1_parabola} - \eqref{eq:eta2_parabola}. --- b) The horizontal distance $\delta h = h_2 - h_1$ between interfaces (I) and (II) exhibits a strong and localized decrease towards its asymptotic value $\delta h_{\infty}$ (black dashed line). --- c) Similarly, the vertical distance $\Delta z$ between interfaces (I) and (II) also plummets, shortly before the two interfaces reach their asymptotic positions in the $x$-direction.}
\label{fig:interfaces}
\end{figure}
Figure~\ref{fig:interfaces}a allows us to appreciate the typical shape of interfaces (I) and (II), as well as their matching to the static menisci (parameters $\Sigma = 0.667$, $R = 0.885$, $M = 1$, $Ca = 10^{-3}$ and $\Delta H = 3.404$).
We focus on scales of the order of $Ca^{2/3}$ and $Ca^{1/3}$, on the horizontal and vertical axes respectively.
The regions displayed correspond to the dynamic menisci and thin liquid films dragged on the plate.
Interfaces (I) (separating liquids 1 and 2) and (II) (between liquid 2 and air) are represented by blue and orange solid lines, respectively.
These are the dynamic solutions obtained by solving equations \eqref{eq:thinfilm_eqns}, but plotted in their dimensionless non-rescaled form $h_i = \hat{h}_i Ca^{2/3}$ so that they may be represented on the same graph as the static menisci.
The blue and orange dashed lines, obtained by solving the static equations \eqref{eq:meniscus_1} and \eqref{eq:meniscus_2} respectively, show the shapes of the static menisci expected when the plate is not moving and the liquids have zero contact angle with it (perfect wetting conditions).
The location of the plate is outlined by the thick vertical grey line at $x = 0$.
In addition to the static menisci themselves, their parabolic expansions, given by \eqref{eq:h1_parabola} and \eqref{eq:h2_parabola}, are displayed with magenta and red dotted lines.
Towards the bath, figure \ref{fig:interfaces}a shows that interfaces (I) and (II) (solid lines) approach these parabolic expansions, as required by the matching conditions \eqref{eq:eta1_parabola} and \eqref{eq:eta2_parabola}.
Using the common terminology in matched asymptotic expansions, this illustrates that the parabolas described by \eqref{eq:h1_parabola} and \eqref{eq:h2_parabola} are simultaneously the inner limits (\textit{i.e.} close to the plate) of the outer solutions (static menisci) and the outer limits (approaching the bath) of the inner solutions (thin liquid films computed using lubrication theory).

Further downstream from the menisci, interfaces (I) and (II) get very close to each other (until they can no longer be told apart at the scale of the plot) and flatten out to converge towards their asymptotic positions.
Figures~\ref{fig:interfaces}b and c allow us to better appreciate this convergence by looking at, on the one hand, the evolution of the horizontal distance $\delta h = h_2 - h_1$ between interfaces (I) and (II) with the vertical coordinate $z$ (panel b) and, on the other hand, the evolution of the vertical distance $\Delta z$ between these interfaces with the horizontal coordinate $x$ (panel c).
Both panels reveal a very abrupt decay of the distance between interfaces (I) and (II) as they approach their asymptotic positions $x=h_{1,\infty}$ and $x=h_{2,\infty}$, respectively.
Asymptotically, the uniform and steady thicknesses bounded by these interfaces are $h_{1,\infty}$ for the film of liquid 1 (referred to as \emph{lower film}) and $\delta h_{\infty} = h_{2,\infty} - h_{1,\infty}$ for the film of liquid 2 (referred to as \emph{upper film}).

Remarkably, the thickness of the upper film is much smaller than that of the lower one: in the example of showed in figure \ref{fig:interfaces}, $\delta h_{\infty} = 1.97 \times 10^{-4}$ while $h_{1,\infty} = 1.33 \times 10^{-2}$ or, in terms of rescaled quantities, $\delta\hat{h}_\infty =\delta h_\infty \, Ca^{-2/3} = 1.97 \times 10^{-2}$ while $\hat{h}_{1,\infty} = h_{1,\infty} \, Ca^{-2/3} = 1.33$. Note that, with this normalization, the rescaled thickness for a one-liquid Landau-Levich-Derjaguin film would simply be $\hat{h}_{\mathrm{LLD}} = 0.9458$, as the corresponding dimensional thickness is equal to $0.9458 \, \ell_c \, Ca^{2/3}$ \citep{LandauLevich1942}.
As we will see further down in section \ref{sec:results2}, this feature ($\delta\hat{h}_\infty \ll \hat{h}_{1,\infty}$) is observed in all the parameter ranges explored in this work.
We can already notice that differences in the properties of liquids 1 and 2 are not sufficient to account for such a discrepancy because the dimensional capillary lengths $\ell_{c,1}$ and $\ell_{c,2}$, corresponding to interfaces (I) and (II) respectively, remain of the same order of magnitude: $\ell_{c,2}/\ell_{c,1} = \sqrt{\Sigma \left(1/R - 1\right)} = 0.208 - 0.406$
for all parameters considered in our study (see table \ref{tab:param_table}).
As we will develop in next paragraphs, the difference in the order of magnitude of $h_{1,\infty}$ and $\delta h_{\infty}$ is connected to the very different flow patterns conveying liquid into the lower and upper films.
%
\subsection{Flow structure} \label{ssec:flow_structure}
%
\begin{figure}
\centering
\includegraphics[width=13cm]{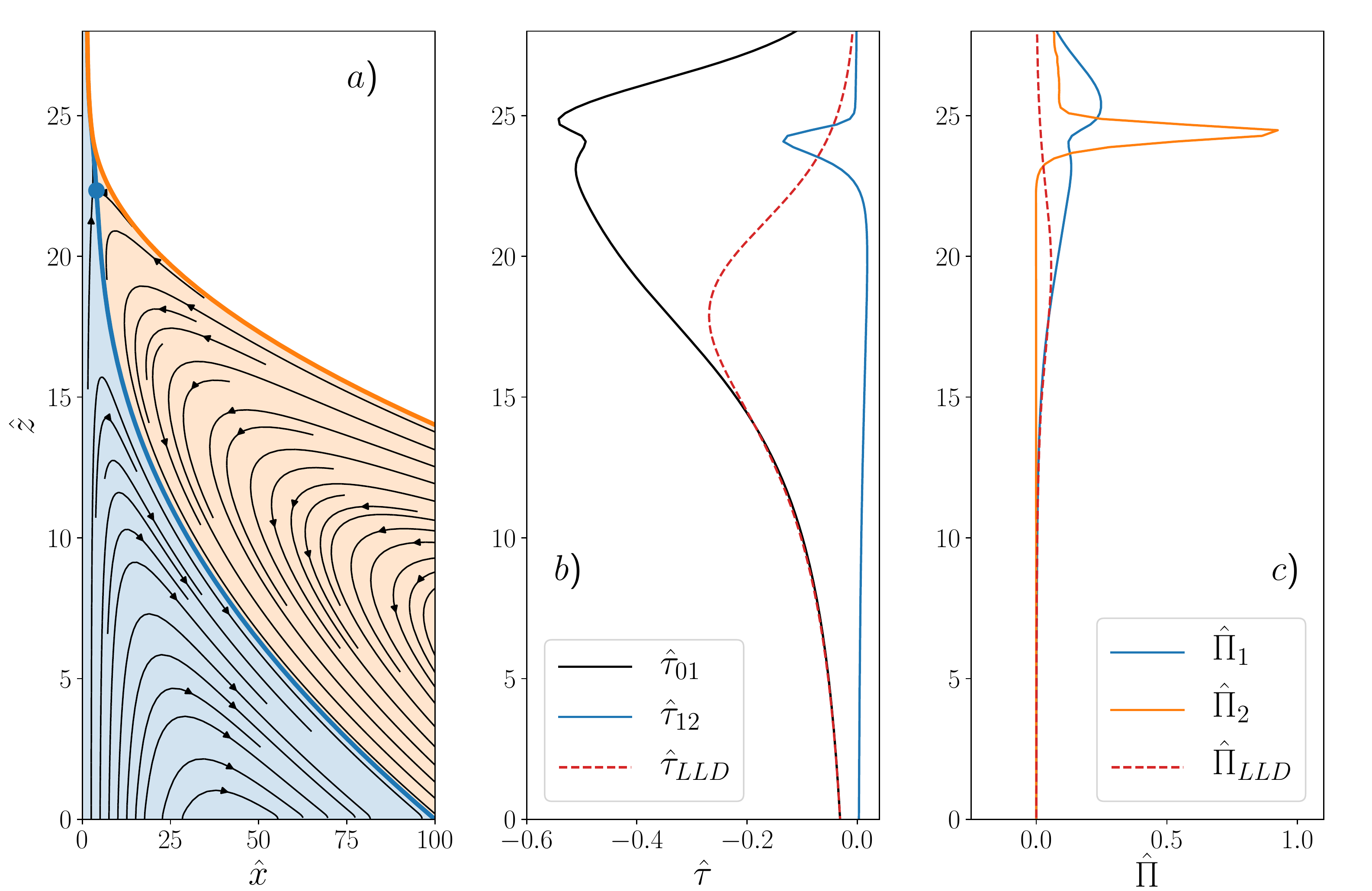}
\caption{Flow structure for dimensionless parameters $\Sigma = 0.667$, $R = 0.885$, $M = 1$, $Ca = 10^{-3}$ and $\Delta H = 3.403$ as a function of the vertical coordinate $\zhat$. (a) Streamlines in liquid 1 (blue) and liquid 2 (orange). --- (b) Shear stresses at the plate/liquid 1 interface, $\htau_{01}$ (black solid line), and at the liquid 1/liquid 2 interface, $\htau_{12}$ (blue solid line). --- (c) Pressure gradients in liquid 1, $\hPi_1$ (solid blue line), and in liquid 2, $\hPi_2$ (solid orange line). For comparison, the red dashed lines represent the corresponding magnitudes in the one-liquid LLD theory: shear stress at the plate/liquid interface, $\htau_{\rm LLD}$ (b), and pressure gradient in the liquid, $\hPi_{\rm LLD}$ (c).}
\label{fig:streamlines_shear_pressure_grad}
\end{figure}
In this paragraph, we turn our attention to the structure of the flow in the dynamic meniscus region.
As an example, figure~\ref{fig:streamlines_shear_pressure_grad} displays various relevant flow magnitudes, obtained for parameters $\Sigma = 0.667$, $R = 0.885$, $M = 1$, $Ca = 10^{-3}$ and $\Delta H = 3.404$, as functions of the vertical coordinate $\zhat$.
In figure~\ref{fig:streamlines_shear_pressure_grad}a, we plot the streamlines in the dynamic menisci for liquid phases 1 (in blue) and 2 (in orange).
The corresponding analytical expressions for the velocity fields are given in Appendix~\ref{apdx:velocity_and_shear} by equations (\ref{eq:u1}--\ref{eq:v1}) for liquid 1 and (\ref{eq:u2}--\ref{eq:v2}) for liquid 2, respectively. 
While, as expected, the plate drags the lower liquid up, streamlines reveal that the flow in the vicinity of interface (I) (liquid 1/liquid 2 interface, thick blue line) is actually going \emph{downwards} upstream the stagnation point (blue dot).
The consequences of this flow pattern on the coated liquid films can be better understood by looking at the main physical effects at play: viscous entrainement by shear stresses at the dragging interfaces (plate/liquid 1 for the lower phase, liquid 1/liquid 2 for the upper phase) and capillary suction generated by corresponding interfacial curvatures.
The former is presented in figure~\ref{fig:streamlines_shear_pressure_grad}b, where the dimensionless rescaled shear stress at different interfaces of interest is plotted as a function of the vertical coordinate $\zhat$, while the latter is quantified in figure~\ref{fig:streamlines_shear_pressure_grad}c, displaying the dimensionless rescaled pressure gradient as a function of $\zhat$.

%
Let us first focus on the viscous stresses, which promote liquid film entrainment.
The solid black line in figure~\ref{fig:streamlines_shear_pressure_grad}b represents the shear stress $\htau_{01}$, defined by equation~\eqref{eq:def_htau0}, at the interface between the plate and liquid 1 in our two-liquid configuration.
For comparison, the dashed red line shows the shear stress $\htau_{LLD}$ at the plate/liquid interface in the one-liquid film coating problem; the corresponding expression is given by equation~\eqref{eq:def_htauLLD}.
Both curves exhibit qualitatively the same shape: the shear stress at the plate increases progressively as the height above the surface of the bath increases, goes through a maximum and then decays back to zero as $\zhat$ keeps increasing.
This is in sharp contrast with the trend followed by the shear stress $\htau_{12}$ at interface (I), separating liquids 1 and 2 (solid blue line, defined by equation~\eqref{eq:def_htau1}).
The shear stress at interface (I) is found to be essentially zero everywhere, except for a small peak in a narrow region (around $\zhat \approx 24$ in this example), which coincides with the zone where interfaces (I) and (II) get very close to each other (see figure~\ref{fig:interfaces} and § \ref{ssec:shape_interfaces}).
This observation is consistent with the streamlines of figure~\ref{fig:streamlines_shear_pressure_grad}a that show that liquid 2 is essentially recirculating on top of liquid 1, as no shear is transmitted along most of interface (I).
Note that the shear stress at interface (II) (separating liquid 2 and the atmosphere) is identically zero, as set by the boundary condition \eqref{eq:BCx_adim_1}.

%
We now turn to the capillary pressure gradients, which impede thin liquid film entrainment.
Figure~\ref{fig:streamlines_shear_pressure_grad}c displays the dimensionless rescaled pressure gradient $\hPi_1$ (blue) and $\hPi_2$ (orange) in liquid phases 1 and 2, defined by equations~\eqref{eq:def_hPi1} and \eqref{eq:def_hPi2}, respectively.
Again, the two liquid phases exhibit very different behaviors: the lower liquid is subjected to a mild pressure gradient, spread over distances of several units in $\zhat$.
On the contrary, there is virtually no pressure gradient in the upper liquid, except for a high peak around $\zhat \approx 24$, which also corresponds to the region of highest shear stress (in absolute value) at the liquid 1/liquid 2 interface.
%
\subsection{Geometrical approach: virtual contact point} \label{ssec:geometrical_model}
%
\begin{figure}
\centering
\vspace{1mm}
\includegraphics[width=8cm]{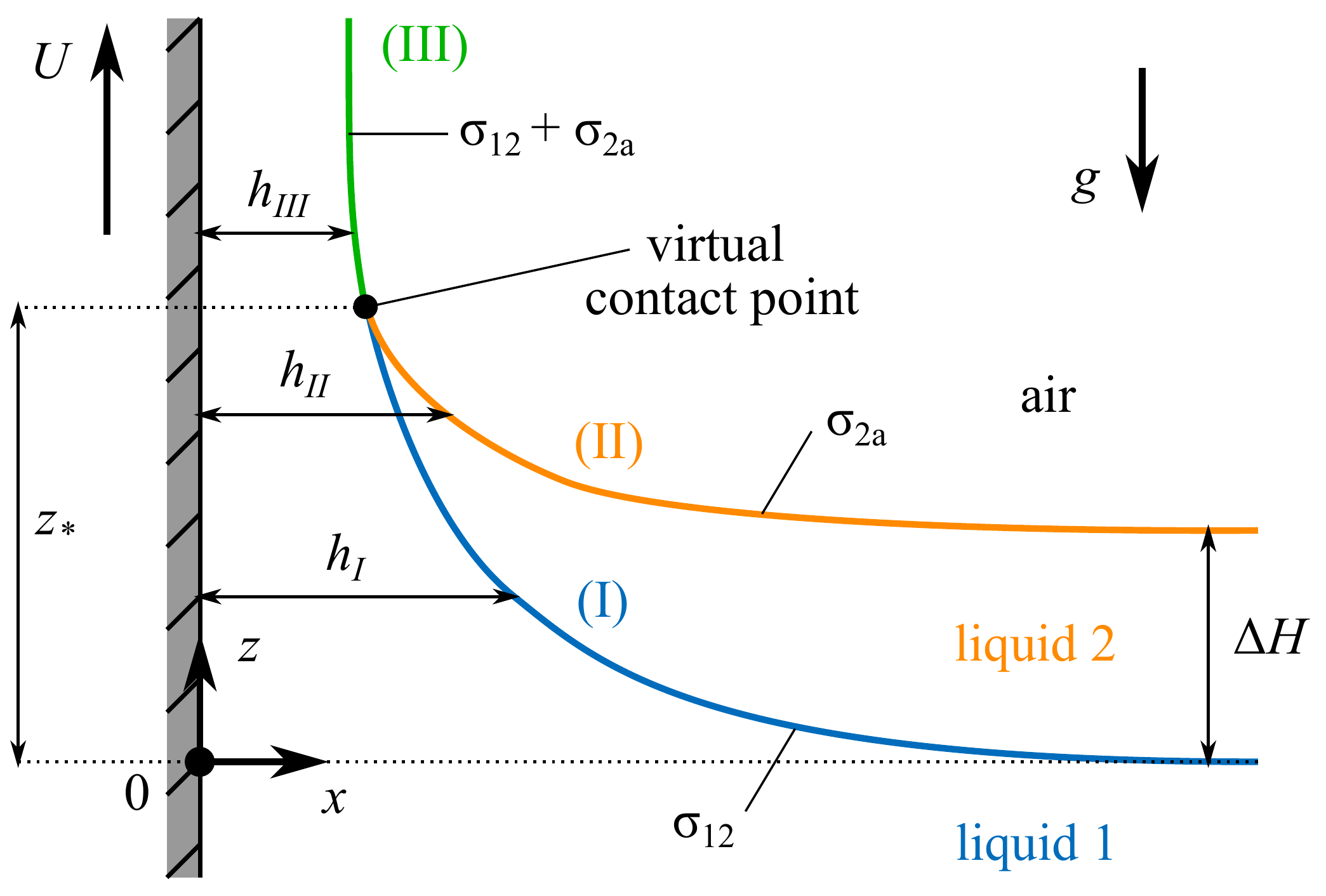}
\caption{Sketch showing the virtual contact point, where interfaces (I) and (II) are assumed to come into contact. Above this point, located at height $z^{\ast}$, interfaces (I) and (II) merge into a single interface, denoted by (III), with an effective surface tension equal to the sum of that of interfaces (I) and (II).}
\label{fig:sketch_contact_point}
\end{figure}
Let us focus on the region where we observe the peak in the shear stress at the liquid/liquid interface, $\htau_{12}$, and in the pressure gradient in the upper film, $\hPi_2$.
Figures \ref{fig:streamlines_shear_pressure_grad}b and \ref{fig:streamlines_shear_pressure_grad}c show that this region is very narrow in the vertical direction, as compared to the overall extension of the dynamic menisci.
Moreover, figure \ref{fig:interfaces}b reveals that, concomitantly, the distance between interfaces (I) and (II) decays quickly down to the asymptotic one, $\delta h_\infty$.
For these reasons, we find meaningful to model this region as a point, called \emph{virtual contact point}, where interfaces (I) and (II) virtually meet.
Note that, in general, this point does not coincide with a stagnation point at interface (I), as can be seen in the example of figure~\ref{fig:streamlines_shear_pressure_grad}.
The vertical coordinate of the virtual contact point is denoted by $z_{\ast}$.
In the following, all quantities evaluated at this point will be marked by an $\ast$.

Introducing the virtual contact point allows us to develop a geometrical, asymptotic (``zoomed-out'') description of the flow, as sketched in figure \ref{fig:sketch_contact_point}.
By construction, we have $\delta h \ll h_{1}$, and asymptotically $\delta h_{\infty} \ll h_{1,\infty}$, downstream the virtual contact point. 
We therefore simplify the geometry in this region, assuming that interfaces (I) and (II) are merged into a single interface (III) with effective dimensionless surface tension $1 + \Sigma$.
In this framework, the different interfaces pictured in figure \ref{fig:sketch_contact_point} can be described as follows. \medskip

\begin{itemize}
    \item Interface (I) --- In the region below the virtual contact point, the very small shear stress acting on interface (I) (see figure \ref{fig:streamlines_shear_pressure_grad}b) implies that this surface can be regarded as shear-free.
    As a consequence, the lower liquid 1 bounded by this interface behaves as a one-liquid Landau-Levich film.
    \item Interface (II) --- The upper liquid 2 is bounded by the approximately shear-free interface (I) and the rigorously shear-free interface (II). Consequently, the pressure gradient inside this liquid  must be zero, which is consistent with figure \ref{fig:streamlines_shear_pressure_grad}c.
    \item Interface (III) --- In the region above the virtual contact point, by construction, the dynamics of the effective interface (III) also obeys the one-liquid Landau-Levich equation.
\end{itemize} \medskip

This simplified description will be useful to explain some of the most salient features of the two-liquid flow using well-known properties of its one-liquid counterpart.
More specifically, we will show that the asymptotic film thicknesses can be rationalized, and even quantitatively predicted to some extent, looking at the flow variables at the virtual contact point.
%
\subsection{Scaling laws for the film thicknesses} \label{ssec:scaling_laws}
%
\begin{figure}
\centering
\includegraphics[trim=0.3cm 0cm 0.7cm 0cm,clip,width=\linewidth]{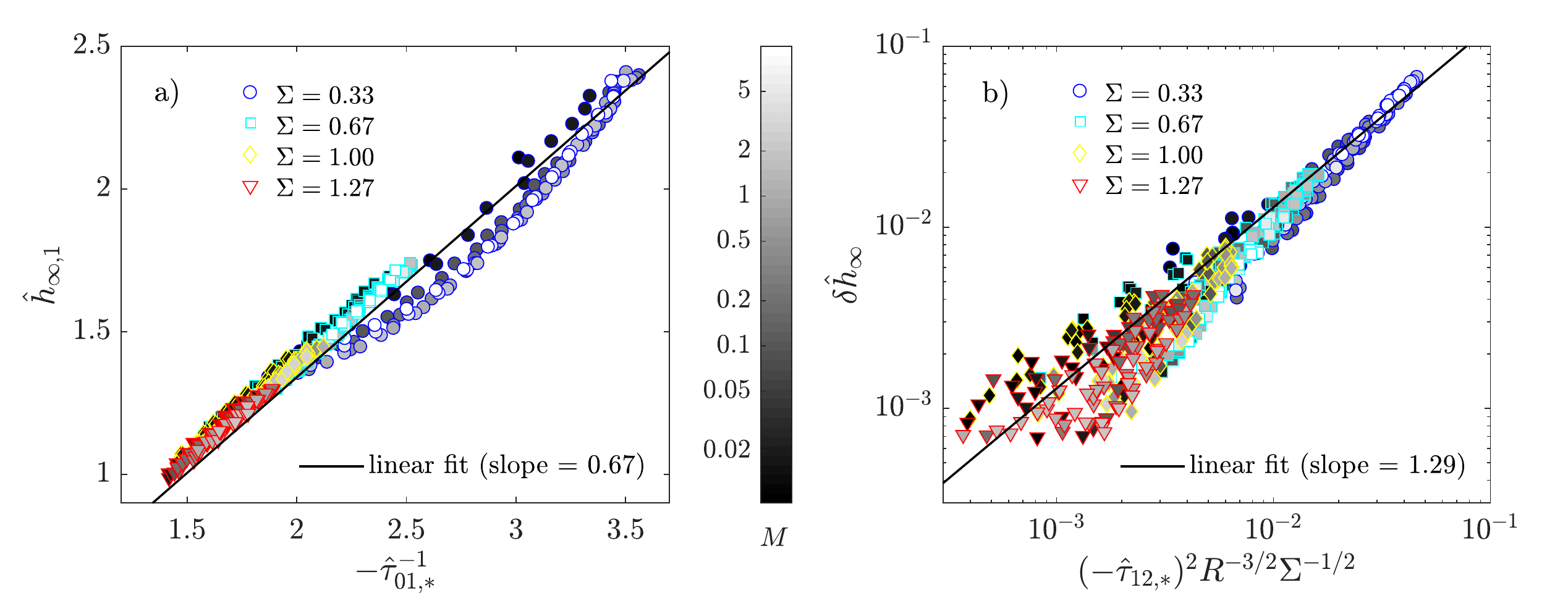}
\caption{Comparison of the asymptotic thicknesses $\hhat_{1,\infty}$ and $\delta\hhat_\infty$ obtained numerically with the predictions of scaling laws derived from the virtual contact point model for a) the lower film (equation \eqref{eq:scaling_h1infty_tau0}) and b) the upper film (equation \eqref{eq:scaling_deltahinfty_tau1}). The solid lines are best linear fits, yielding prefactors $0.67$ and $1.29$ for the lower and upper films, respectively. The data presented corresponds to various values of $\Sigma$ (symbol shape), $M$ (greyscale) and $\Delta H$ (not marked).}
\label{fig:plot_vs_tau}
\end{figure}
The results presented in figure \ref{fig:streamlines_shear_pressure_grad} show that the two main competing forces -- viscous stresses and capillary pressure gradient -- reach their extreme values in the vicinity or at the virtual contact point.
This suggests that the amount of liquid entrained in each film may be rationalized by considering solely the local shear stresses around that location.
In what follows, we use scaling arguments -- in the spirit of the ones developed, for instance, in \cite{Champougny2015} -- to relate the (dimensionless rescaled) shear stresses at the virtual contact point to the steady-state (dimensionless rescaled) thicknesses of the coated liquid films, namely $\hat{h}_{1,\infty}$ for liquid 1 and $\delta\hat{h}_\infty$ for liquid 2.
%
\subsubsection*{Lower film thickness $\hhat_{1,\infty}$}
%
In the case of the lower film, the shear stress responsible for liquid entrainment is the one at the plate/liquid 1 interface (black solid line in figure \ref{fig:streamlines_shear_pressure_grad}b).
In dimensional terms, the maximum shear stress must be of the order of the viscosity times the plate velocity divided by the minimum thickness, \textit{i.e.} that of the film:
\begin{equation}
    -\tau_{01, \ast} \sim \frac{\mu_1 U}{h_{1,\ast}}.
\end{equation}
In this expression, the star indicates that the quantities are evaluated at the virtual contact point.
The dimensional thickness and stress are related to their dimensionless rescaled counterparts through $h_{1,\ast} = \hhat_{1,\ast} \, \ell_c \, Ca^{2/3}$ and $\tau_{01, \ast} = \hat{\tau}_{01,\ast} \, (\sigma_{12}/\ell_c) \, Ca^{1/3} $, respectively, yielding $-\hat{\tau}_{01,\ast} \hhat_{1,\ast} \sim 1$.
Making the approximation that the film thickness has already reached its asymptotic value at the virtual contact point, namely that $\hhat_{1,\infty} \approx \hhat_{1,\ast} $, we arrive at
\begin{equation}
    \hhat_{1,\infty} \sim -\hat{\tau}_{01,\ast}^{-1}.
    \label{eq:scaling_h1infty_tau0}
\end{equation}
In figure \ref{fig:plot_vs_tau}a, we plot the lower film thickness $\hhat_{1,\infty}$ as a function of the inverse of the shear stress at the virtual contact point, $-\hat{\tau}_{01,\ast}^{-1}$, for a variety of $\Sigma$, $M$ and $\Delta H$. 
All numerical data are found to collapse on a single master curve, which is convincingly adjusted by a linear fit (solid black line) with a prefactor equal to $0.67$, therefore supporting the scaling law \eqref{eq:scaling_h1infty_tau0}.
Note that the presence of a second lighter fluid always causes the lower film thickness $\hhat_{1,\infty}$ to be larger than that of a one-liquid Landau-Levich film $\hhat_{\rm LLD} = 0.9458$.
%
\subsubsection*{Upper film thickness $\delta \hhat_{\infty}$}
%
In the case of the upper film, the shear stress responsible for liquid entrainment is the one at interface (I), \textit{i.e.} at the liquid 1/liquid 2 interface (blue solid line in figure \ref{fig:streamlines_shear_pressure_grad}b).
Unlike the one at the plate, the velocity at interface (I) is not given \textit{a priori} but depends in a non-trivial way on the parameters of the problem.
For this reason, we cannot start from the definition of the shear stress, as we did in the case of the lower film.
Instead, we write that, around the virtual contact point, the total shear force at interface (I) must balance the capillary suction exerted by interface (II).
Denoting by $\ell$ the streamwise extension of the virtual contact point region (typically the width of the peaks in $\hat{\tau}_{12}$ and $\hat{\Pi}_2$ in figure \ref{fig:streamlines_shear_pressure_grad}), this force balance reads in dimensional terms:
\begin{equation}
    -\tau_{12}^{\ast} \, \ell \sim \sigma_{2a} \frac{\delta h_{\ast}}{\ell^2} ~ \delta h_{\ast}.
    \label{eq:scaling_force_balance_II}
\end{equation}
Additionally, matching the curvatures of the dynamic and static menisci of liquid 2 imposes (still using dimensional quantities)
\begin{equation}
    \frac{\delta h_{\ast}}{\ell^2} \sim \frac{1}{\ell_{c,2}},
    \label{eq:scaling_static_meniscus_matching_II}
\end{equation}
where we recall that $\ell_{c,2} = \sqrt{\sigma_{2a}/\rho_2 g}$ is the capillary length related to interface (II).
This condition is analogous to the asymptotic matching condition with the static meniscus introduced by \citet{LandauLevich1942} in the one-liquid dip-coating problem.
Combining equations \eqref{eq:scaling_force_balance_II} and \eqref{eq:scaling_static_meniscus_matching_II}, we arrive at the dimensional scaling expression
\begin{equation}
    \delta h_{\infty} \sim \ell_{c,2} \, \left(-\tau_{12,\ast} \, \frac{ \ell_{c,2}}{\sigma_{2a}} \right)^2,
    \label{eq:scaling_deltahinfty_dim}
\end{equation}
where we approximated $\delta h_{\infty} \approx \delta h_{\ast}$.
Using the definitions $\delta h_{\infty} = \delta \hhat_{\infty} \, \ell_c \, Ca^{2/3}$ and $\tau_{12, \ast} = \hat{\tau}_{12,\ast} \, (\sigma_{12}/\ell_c) \, Ca^{1/3}$, we finally express equation \eqref{eq:scaling_deltahinfty_dim} in terms of the rescaled dimensionless variables:
\begin{equation}
    \delta\hhat_\infty \sim \left(-\hat{\tau}_{12,\ast} \right)^2 R^{-3/2} \Sigma^{-1/2}.
    \label{eq:scaling_deltahinfty_tau1}
\end{equation}
In figure \ref{fig:plot_vs_tau}b, we plot the upper film thickness $\delta \hhat_{\infty}$ as a function of the right-hand term of equation \eqref{eq:scaling_deltahinfty_tau1} for a variety of $\Sigma$, $M$ and $\Delta H$. 
When plotted in that way, the numerical data is found to collapse on a reasonably linear master curve.
The linear fit (solid black line) yields a prefactor equal to $1.29$, showing that equation \eqref{eq:scaling_deltahinfty_tau1} can be used to estimate the upper film thickness $\delta \hhat_{\infty}$.

To conclude, perhaps the most important lesson learnt from scalings \eqref{eq:scaling_h1infty_tau0} and \eqref{eq:scaling_deltahinfty_tau1} is the universality of the entrainment mechanism. 
Given the values of the shear stresses at the virtual contact point ($\hat{\tau}_{01,\ast}$ for liquid 1, $\hat{\tau}_{12,\ast}$ for liquid 2), the amount of fluid dragged in each film can be readily estimated using the same ideas exposed in the original work of \citet{LandauLevich1942}.
%
\section{Results: parametric dependence of film thicknesses}
\label{sec:results2}
%
In this section, we turn our attention to the parametric dependence of the asymptotic thicknesses of the coated liquid films, $\hhat_{1,\infty}$ for the lower film and $\delta \hhat_{\infty}$ for the upper film.
The main control parameters varied in our study are the dimensionless floating layer thickness $\Delta H$, the viscosity ratio $M$ and the surface tension ratio $\Sigma$ (see § \ref{ssec:control_params}).
Importantly, our results reveal that double coating solutions only exist in finite areas of the parameter space, which we dub {\em existence islands} (§ \ref{ssec:thickness_maps}). We will propose arguments explaining some trends and boundaries observed for the film thicknesses as a function of $\Delta H$ (§ \ref{ssec:min_DeltaH} and § \ref{ssec:trend_with_DeltaH}).
%
\subsection{Thickness maps in the $M - \Delta H$ parameter space} \label{ssec:thickness_maps}
%
\begin{figure}
\centering
\includegraphics[width=\linewidth]{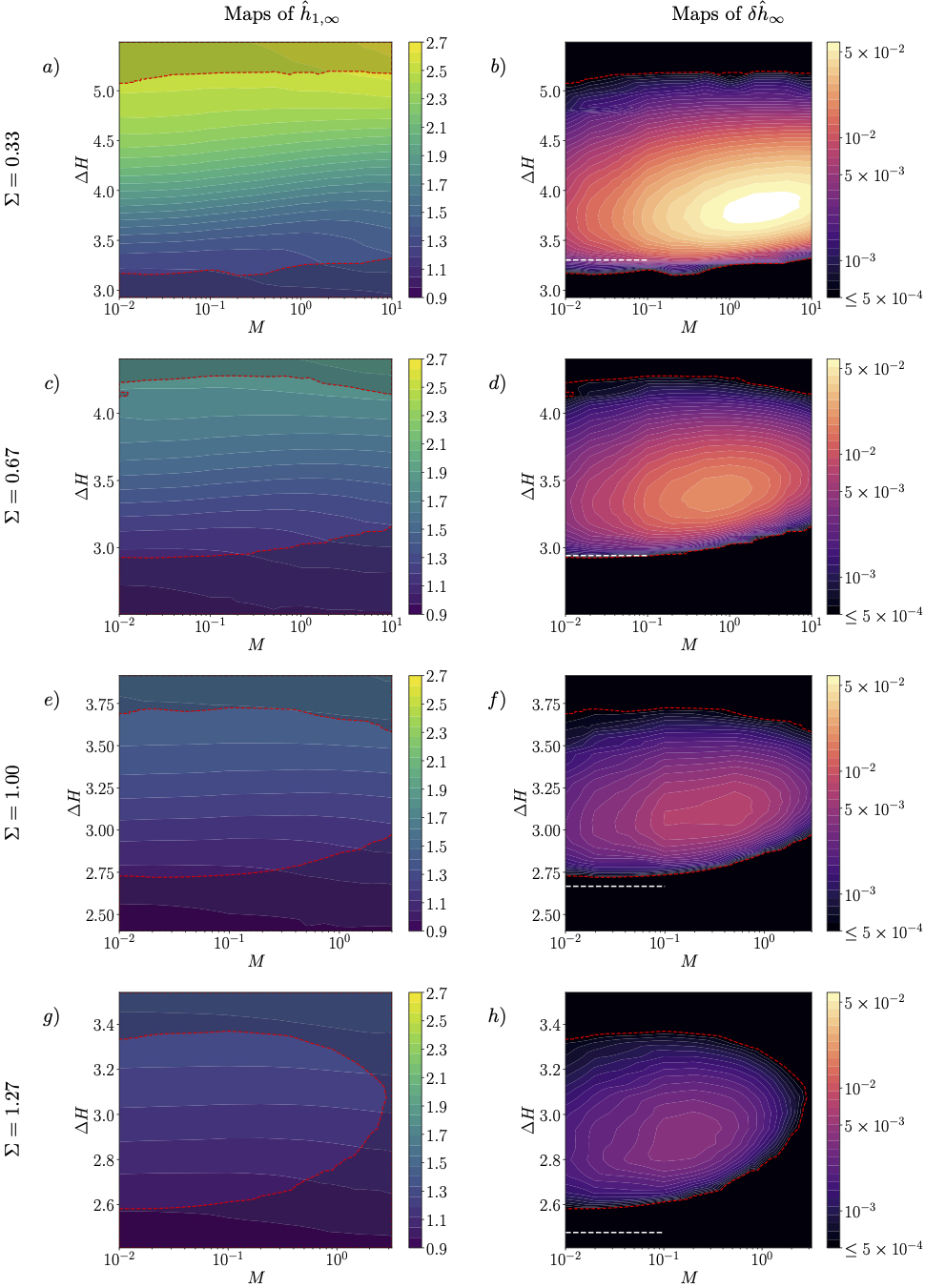}
\caption{Asymptotic thicknesses $\hat{h}_{1,\infty}$ (lower film, in left column) and $\delta\hat{h}_\infty$ (upper film, in right column), shown as color contour maps in the ($M$, $\Delta H$) parameter space. Each row corresponds to a different surface tension ratio $\Sigma$. The red dashed lines enclose the ``existence islands'' of a double coating, namely the areas in the parameter space where $\delta\hat{h}_\infty \ge 5 \times 10^{-4}$. The values of $\hat{h}_{1,\infty}$ obtained outside these islands are shaded.}
\label{fig:thickness_maps}
\end{figure}
\begin{figure}
\centering
\includegraphics[trim=2.5cm 1.5cm 2.5cm 1.5cm,clip,width=0.9\linewidth]{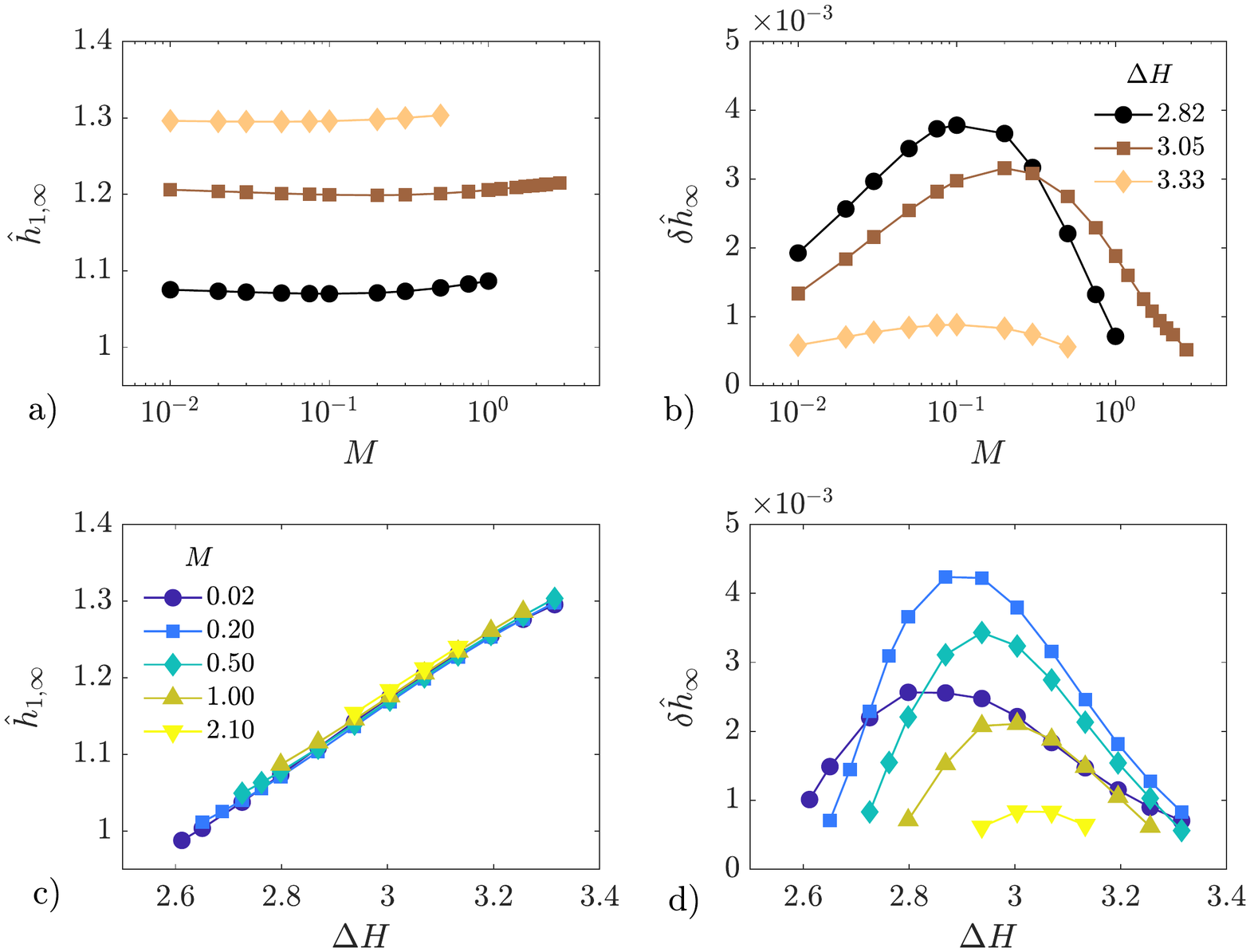}
\caption{Cuts of the thickness maps presented in figures~\ref{fig:thickness_maps}g and \ref{fig:thickness_maps}h ($\Sigma = 1.27$), within the existence island enclosed in the dashed red contour. The top panel shows the variation of (a) the lower film thickness $\hat{h}_{1,\infty}$ and (b) the upper film thicknesses $\delta\hat{h}_\infty$ with the viscosity ratio $M$, for constant values of the floating layer thickness $\Delta H$. The bottom panel shows the variation of the same quantities (on panel (c) and (d), respectively) with the floating layer thickness $\Delta H$, for constant values of the viscosity ratio $M$.}
\label{fig:thickness_cuts}
\end{figure}
In figure~\ref{fig:thickness_maps}, the asymptotic film thicknesses are presented as color maps in the $M - \Delta H$ parameter space for four different values of the surface tension ratio $\Sigma$.
The left column (panels a, c, e, g) shows the \emph{lower} film thickness $\hat{h}_{1,\infty}$ while the right column (panels b, d, f, h) displays the \emph{upper} film thickness $\delta \hat{h}_{\infty}$.
For an easier quantitative reading, cuts along the $M$ and $\Delta H$ directions for the surface tension ratio $\Sigma = 1.27$ are presented in figure~\ref{fig:thickness_cuts}.
The left-hand panels of this figure show the variation of the lower film thickness $\hat{h}_{1,\infty}$ with $M$ for fixed values of $\Delta H$ (figure~\ref{fig:thickness_cuts}a) and with $\Delta H$ for fixed values of $M$ (figure~\ref{fig:thickness_cuts}c).
Similarly, the right-hand panels display the variation of the upper film thickness $\delta \hat{h}_{\infty}$ with the same parameters $M$ (figure~\ref{fig:thickness_cuts}b) and $\Delta H$ (figure~\ref{fig:thickness_cuts}d).

Together, figures~\ref{fig:thickness_maps} and \ref{fig:thickness_cuts} reveal dramatic qualitative and quantitative differences between the lower and upper coated films.
As already observed and rationalized in section \ref{sec:results1}, the lower film exhibits final thicknesses $\hat{h}_{1,\infty}$ of order unity (as expected from the Landau-Levich scaling, see § \ref{ssec:scaling_laws}), while the upper film reaches steady-state thicknesses $\delta\hat{h}_\infty$ that are about $10^{-3}$ to $10^{-2}$ times smaller.
Not only the lower and upper film thicknesses are very disparate, but they also depend very differently on the control parameters.
Figures~\ref{fig:thickness_maps}a, c, e, g reveal that the thickness of the lower film weakly depends on the viscosity ratio $M$ and grows monotonically with the depth of the floating layer $\Delta H$.
On the contrary, figures~\ref{fig:thickness_maps}b, d, f, h show that the thickness $\delta\hat{h}_\infty$ of the upper film depends non-monotonically on both $M$ and $\Delta H$, exhibiting a maximum whose exact position in the ($M, \Delta H$) parameter space depends on the surface tension ratio $\Sigma$.

Around this maximum, the upper film thickness decreases in all directions, eventually going down to values reaching our numerical accuracy ($\delta\hat{h}_\infty \approx 5 \times 10^{-4}$, dashed red lines in figure~\ref{fig:thickness_maps}).
We observed that increasing the resolution of our numerical scheme barely affects the position of the red contours, leading us to conclude that there is no solution with two entrained films beyond these limits (black areas in figures~\ref{fig:thickness_maps}b, d, f, h).
The zones of the ($M, \Delta H$) parameter space enclosed within the red dashed contours will therefore be referred to as \emph{existence islands} for the double-layer coating.
Note that values of the lower film thickness $\hat{h}_{1,\infty}$ can still be obtained outside these islands (see shaded areas in  figures~\ref{fig:thickness_maps}a, c, e, g).
However, since the theory used to obtain them postulates the presence of two entrained liquid films, only the data enclosed in the existence islands -- where a double coating solution exists -- can be discussed.
%
\subsection{Minimum and maximum $\Delta H$ for the existence of two films} \label{ssec:min_DeltaH}
%
The very different scales on the $M$ and $\Delta H$ axes in figure~\ref{fig:thickness_maps} highlight that the existence islands extend only along a finite range of values of $\Delta H$, getting narrower and narrower as $\Sigma$ increases, while they span several orders of magnitude in $M$ values. 
In this paragraph, we aim at providing some physical arguments to rationalize this observation.

The existence islands exhibit a relatively well-defined lower boundary in $\Delta H$, which becomes independent of the viscosity ratio $M$ for $M \ll 1$.
This limit can be understood from hydrostatics considerations.
A necessary condition to have two entrained films is that the static menisci of the two liquids touch the plate, as depicted in figure \ref{fig:sketch_notations_static}. 
In other words, that the apparent contact line of the upper interface lays above that of the lower one. 
For this to occur, the distance between the apparent contact lines of the lower and the upper menisci, $\Delta z_{cl}$, must be positive. Combining equations (\ref{eq:kappa_stat_1}) and (\ref{eq:kappa_stat_2}) this condition translates into
\begin{equation}
    \Delta H \ge \Delta H_{\mathrm{min}} = \sqrt{\frac{2}{1-R}} - \sqrt{\frac{2\Sigma}{R}}.
    \label{eq:DeltaH_min}
\end{equation}
The values of $\Delta H_{\mathrm{min}}$ predicted by equation \eqref{eq:DeltaH_min} for surface tension ratios $\Sigma =$ $0.33$, $0.67$, $1.00$ and $1.27$ are presented in table \ref{tab:discussion_table} and displayed in figures~\ref{fig:thickness_maps}b, d, f, h as dashed white lines.
Comparison with the numerical data reveals a good agreement with the lower limit of the existence islands for $M \ll 1$. 

Regarding the maximum $\Delta H$ for the existence of a double-film configuration, $\Delta H_{\mathrm{max}}$, the physical origin is different. For the upper film to be dragged, shear must be transmitted from the plate to the interface between the two liquids. However, the region where the plate exerts shear on the lower film is limited to an extension of order $\Delta\hat{z} \sim 10$ in the streamwise direction, as can be seen in figure \ref{fig:streamlines_shear_pressure_grad}b. This means that if the virtual contact point -- where the two interfaces come close to each other -- is outside this limited region in $\hat{z}$, no shear can be transferred to the liquid / liquid interface, and thus the upper film cannot be dragged. In terms of $\Delta H$, equation \eqref{eq:relationship_DeltaH_Ca} allows us to translate the extension $\Delta\hat{z} \sim 10$ into 
\begin{equation}
    \Delta H_{\mathrm{max}} - \Delta H_{\mathrm{min}} \sim 10 \, Ca^{1/3}.
        \label{eq:DeltaH_max}
\end{equation}
The condition $Ca^{1/3} \ll 1$ (see § \ref{ssec:rescaling} and § \ref{ssec:validity_conditions}) explains why we expect the existence islands to span only along a reduced extension in $\Delta H$. In our particular case where $Ca = 10^{-3}$, equation \eqref{eq:DeltaH_max} predicts island extensions of the order of unity in $\Delta H$, which is compatible with the data displayed in figure \ref{fig:thickness_maps}.
%
\subsection{Effect of $\Delta H$ on the lower film thickness $\hhat_{1,\infty}$} \label{ssec:trend_with_DeltaH}
%
\begin{table}
    \centering
    \begin{tabular}{ccccccc}
      & Quantity & $\quad \Sigma=0.33 \quad$ & $\Sigma=0.67$ & $\quad\Sigma=1.00$ &%
      $\quad \Sigma=1.27 \quad$ & Source\\
    \hline\hline
        (a) & Lowest $\Delta H$ & $3.16$ & $2.94$ & $2.72$ & $2.57$ & Fig.~\ref{fig:thickness_maps}\\
        & $\Delta H_{\mathrm{min}}$ & $3.30$ & $2.94$ & $2.67$ & $2.48$ & Eq.~\eqref{eq:DeltaH_min} \\
        \hline
        (b) & $\mathrm{min} \, (\hhat_{1,\infty})$ & $1.27$ & $1.09$ & $1.01$ & $0.97$ & Fig.~\ref{fig:thickness_maps} \\
        & $\mathrm{inf} \, (\hhat_{1,\infty})$ & $1.18$ & $1.01$ & $0.93$ & $0.89$ & Eq.~\eqref{eq:lower_bound_h1infty}\\
        \hline
    \end{tabular}
    \caption{Comparison between some features of the results presented in figure \ref{fig:thickness_maps} and the corresponding values predicted using simplified approaches (section \ref{sec:discussion}), for different values of the surface tension ratio $\Sigma$. The quantities extracted from figure \ref{fig:thickness_maps} are evaluated in the limit $M \ll 1$ and within the existence islands (red dashed contours), where a double coating solution exists. These quantities are (a) the lower limit in $\Delta H$ of the existence islands and (b) the minimum values of the asymptotic lower film thickness $\hhat_{1,\infty}$.}
    \label{tab:discussion_table}
\end{table}
In this paragraph we use the equivalent description presented in § \ref{ssec:geometrical_model} to explain the effect of the floating layer thickness $\Delta H$ on the asymptotic thickness $\hhat_{1,\infty} \approx \hhat_{III,\infty}$ of the lower film (see figure \ref{fig:sketch_contact_point}). 
To do so, we first examine the curvatures of the different interfaces represented in figure \ref{fig:sketch_contact_point} and make the following observations. \medskip

\begin{itemize}
    \item Interface (I) --- Since interface (I) behaves as in the one-liquid LLD theory, its curvature $\hhat''_{I}$ decays monotonically with $\hat{z}$ (see appendix \ref{apdx:one_liquid_LLD}).
    The maximum curvature, $\sqrt{2\left(1-R\right)}$, is found in the limit $\zhat \rightarrow -\infty$, corresponding to the lower static meniscus.
    \item Interface (II) --- Since the pressure gradient in liquid 2 is approximately zero, interface (II) has a constant curvature, given by that of the upper static meniscus: $\hhat''_{II} =\sqrt{2R/\Sigma}$.
    \item Interface (III) --- For the same reason as interface (I), the curvature of interface (III) decreases monotonically with $\hat{z}$, reaching its maximum value $\hhat''_{III, \ast}$ at its lowest point, \textit{i.e.} the virtual contact point. 
\end{itemize} \medskip

Taking again advantage of the fact that interface (III) can be described by the one-liquid LLD theory, we approximate (see equation \eqref{eq:h_infty_vs_K_LLD})
\begin{equation}
    \hhat_{III,\infty} \approx \frac{1.336}{(1+\Sigma)^{2/3} \, \hhat''_{III, \ast}}.
    \label{eq:approx_curvature_intIII}
\end{equation}
Moreover, the condition that the pressure inside liquid 1 must be continuous across the virtual contact point determines the value of the curvature of interface (III) at this location. On the one hand, the pressure slightly downstream the virtual contact point is given by the capillary pressure jump across interface (III), $(1 + \Sigma) \, \hat{h}''_{III,*}$. On the other hand, just upstream that point, the pressure is equal to the sum of the capillary pressure jumps across interfaces (II) and (I), which are respectively $\Sigma\hat{h}''_{II,*}$ and $\hat{h}''_{I,*}$. Requiring pressure continuity at the virtual contact point leads to
\begin{equation}
    \hhat''_{III, \ast} = \frac{\hhat''_{I, \ast} + \Sigma \hhat''_{II}}{1+\Sigma}.
    \label{eq:pressure_cont_condition}
\end{equation}
Combining equations \eqref{eq:approx_curvature_intIII} and \eqref{eq:pressure_cont_condition}, we arrive at the following expression, relating the asymptotic lower film thickness to the curvature of interface (I) at the virtual contact point:
\begin{equation}
    \hhat_{1,\infty} \approx \hhat_{III,\infty} \approx %
    \frac{1.336 \, (1+\Sigma)^{1/3}}{\hhat''_{I, \ast} + \sqrt{2\Sigma R}}.
    \label{eq:h_III_intfy_vs_curv_of_interface_I_star}
\end{equation}
%
\subsubsection*{Monotonous behaviour of $\hhat_{1,\infty}$ with $\Delta H$}
%
This simplified view allows us to explain why $\hhat_{1,\infty}$ grows monotonically with the thickness of the floating layer, $\Delta H$. 
As this parameter grows, the virtual contact point, where interfaces (I) and (II) meet, displaces up downstream or, in other words, $\hzstar$ increases.
Since $\hhat''_{I}$ is a decreasing function of $\hat{z}$, the larger $\hzstar$, the lower the corresponding curvature $\hhat''_{I, \ast}$ of interface (I).
We therefore deduce that $\hhat''_{I, \ast}$ decays with $\Delta H$.
Finally, equation \eqref{eq:h_III_intfy_vs_curv_of_interface_I_star} allows us to conclude that $\hhat_{1,\infty}$ must be a monotonously increasing function of $\Delta H$, as observed in our numerical results in figures \ref{fig:thickness_maps} and \ref{fig:thickness_cuts}.
%
\subsubsection*{Lower bound for $\hhat_{1,\infty}$}
%
Elaborating on these ideas, we can also provide a lower bound for the asymptotic thickness of the lower film. 
If the virtual contact point is displaced far upstream ($\zhat \rightarrow -\infty$, \textit{i.e.} closer to the liquid bath), it will eventually reach the region where the curvature of interface (I) has its asymptotic (maximum) value: $\hhat''_{I, \ast} = \sqrt{2\left(1-R\right)}$. 
Substituting this value in equation \eqref{eq:h_III_intfy_vs_curv_of_interface_I_star} yields a lower bound for the lower film thickness:
\begin{equation}
    \mathrm{inf} \, (\hhat_{1,\infty}) \approx 0.9458 \frac{\left(1+\Sigma\right)^{1/3}}{\sqrt{1-R} + \sqrt{\Sigma R}}.
    \label{eq:lower_bound_h1infty}
\end{equation}
We can compare the predictions of equation \eqref{eq:lower_bound_h1infty} to the values of $\hhat_{1,\infty}$ observed in figure \ref{fig:thickness_maps}.
To do so, we should restrict ourselves to the area of the parameter space enclosed in the dashed red line (where solutions for two entrained films exist) and to the limit $M \ll 1$, in which the simplified geometrical model is valid.
As shown in table \ref{tab:discussion_table}, the minimum values $\mathrm{min} \, (\hhat_{1,\infty})$ obtained from the simulations are not only compatible with the lower bounds provided by equation \eqref{eq:lower_bound_h1infty} but also follow a similar trend with $\Sigma$.
%
\section{Discussion} \label{sec:discussion}
%
In this last section, we discuss some limitations of our hydrodynamic model in relation to practically relevant effects, such as partial wetting conditions betweeen the two liquids and long range intermolecular forces across the upper film, when it is sufficiently thin.
%
\subsection{Partial wetting conditions}
\label{ssec:partial_wetting}
%
As mentioned in § \ref{ssec:validity_conditions}, the stability of a floating layer of liquid 2 on liquid 1 depends on the dimensionless spreading parameter $\mathscr{S} = S/\sigma_{12} = \Sigma^{\prime} - (\Sigma + 1)$, where we have introduced $\Sigma^{\prime} = \sigma_{1a}/\sigma_{12}$. 
For $\mathscr{S} > 0$, the floating layer is stable regardless of its thickness while, for $\mathscr{S} < 0$, only floating layers of dimensionless thickness $\Delta H > \Delta H_c$ given by equation \eqref{eq:floating_layer_DeltaHc} are stable \citep{DeGennes2013}.
In § \ref{ssec:min_DeltaH}, we estimated the minimum floating layer thickness $\Delta H_{\mathrm{min}}$ needed for a double liquid film to be entrained (equation \eqref{eq:DeltaH_min}). 
In partial wetting conditions, our hydrodynamic description of the configuration of figure \ref{fig:sketch_coordinates_definitions} is therefore warranted as long as $\Delta H_{\mathrm{min}} > \Delta H_c$. 
This translates into the following condition on the dimensionless spreading parameter 
\begin{equation}
     - \left( \sqrt{R} - \sqrt{\Sigma (1-R)} \right)^2 < \mathscr{S} < 0, 
\end{equation}
or, equivalently, on the dimensionless liquid 1 / air surface tension $\Sigma^{\prime}$
\begin{equation}
     1 + \Sigma - \left( \sqrt{R} - \sqrt{\Sigma (1-R)} \right)^2 < \Sigma^{\prime} < 1 + \Sigma. 
\end{equation}
%
\subsection{Thinness of the upper film and disjoining pressure effects}
\label{ssec:disjoining_pressure}
%
As can be seen in figure \ref{fig:thickness_maps}, the dimensionless rescaled thickness of the upper film, $\delta\hhat_\infty$, is of the order of $10^{-3}$ to $10^{-2}$ in the range of parameters explored.
The corresponding dimensional thickness, $\delta h_\infty = \ell_c Ca^{2/3} \times \delta\hhat_\infty$, is therefore expected to be in the range $20 - 200~\nano\meter$ for the typical values $Ca = 10^{-3}$ and $\ell_c \approx 2~\milli\meter$.
Albeit small, these thicknesses are withing reach of techniques such as reflection interference contrast microscopy (RICM), as examplified by the recent measurements of \citet{Kreder2018} on the oil layer wrapping water droplets advancing on lubricant-infused surfaces.

Given the thinness of the upper liquid film, one may wonder in which conditions intermolecular long-range forces would be expected to have a noticeable effect on the upper film dynamics. 
These interactions in the film are usually quantified by the disjoining pressure isotherm \citep{Derjaguin1978}, which measures the relative force acting between its two interfaces as a function of their separation.
For asymmetric films made of a pure liquid, such as our upper film of liquid 2, the disjoining pressure isotherm only contains van der Waals interactions, whose nature (attractive or repulsive) depends on the sign of the Hamaker constant $A$ \citep{Israelachvili2011}.
Following the approach of \cite{Champougny2017}, the effect of van der Waals interactions can be included in our formulation by adding a disjoining pressure gradient $\hat{\Pi}^{\rm d}$ to the capillary pressure gradients $\hat{\Pi}_1$ and $\hat{\Pi}_2$ defined in Appendix \ref{apdx:velocity_and_shear}. 
Considering only non-retarded van der Waals forces \citep{Leger1992} and using the rescaling exposed in § \ref{ssec:non_dimensionalization} and § \ref{ssec:rescaling}, the disjoining pressure gradient reads
\begin{equation}
    \hat{\Pi}^{\rm d} = Ca^{-2} \mathcal{A} \frac{\hat{h}'_2 - \hat{h}'_1}{(\hat{h}_2-\hat{h}_1)^4},
\end{equation}
where $\mathcal{A} = A / 2\pi \sigma_{12} \ell_c^2$ is a dimensionless Hamaker constant. 
Using the values of $Ca$, $\sigma_{12}$ and $\ell_c$ reported in § \ref{ssec:control_params}, as well as the typical dimensional Hamaker constant $|A| \approx 10^{-20}~\joule$ for water-oil-air systems \citep{Israelachvili2011}, we estimate $\mathcal{A} \approx 1.6 \times 10^{-14}$.\\
The upper film thickness $\delta\hat{h}_\infty^{\rm crit}$ at which the disjoining pressure gradient becomes of the order of the capillary one can be estimated by requiring:
\begin{equation}
    \hat{\Pi}_2 = -\Sigma \hat{h}'''_2 \sim \hat{\Pi}^{\rm d} = Ca^{-2} \mathcal{A} \frac{\hat{h}'_2 - \hat{h}'_1}{(\hat{h}_2-\hat{h}_1)^4}.
    \label{eq:condition_disjoining_pressure_grad}
\end{equation}
In the rescaled variables used here, $\hat{h}_{2,\infty} \sim \hat{h}_{1,\infty}$ is of order unity, and so is the characteristic distance $\Delta\hat{z}$ over which the thicknesses vary. 
Thus, we have $\hat{h}'''_2 \sim 1$ and $\hat{h}'_2-\hat{h}'_1 \sim \hat{h}_2-\hat{h}_1 = \delta\hat{h}_\infty$, yielding
\begin{equation}
    \delta\hat{h}_\infty^{\rm crit} \sim \left(|\mathcal{A}| \, Ca^{-2} \, \Sigma^{-1}\right)^{1/3}.
\end{equation}
With the values of $\mathcal{A}$ and $Ca$ given above, we arrive to thresholds $\delta\hat{h}_\infty^{\rm crit} \sim 3.6 \times 10^{-3}$, $2.9 \times 10^{-3}$, $2.5 \times 10^{-3}$ and $2.3 \times 10^{-3}$ for $\Sigma = 0.333$, $0.667$, $1.00$ and $1.27$, respectively. 
For each $\Sigma$, this means that the parts of the existence islands where $\delta\hat{h}_\infty \lesssim \delta\hat{h}_\infty^{\rm crit}$ (essentially the edges, see figure \ref{fig:thickness_maps}) may be affected by disjoining pressure.
Conversly, disjoining pressure effects are not expected to play a significant role in the interior of the existence islands.
Note that the above reasoning allows us to estimate when van der Waals interactions are expected to influence the value of the asymptotic \emph{thickness} of the upper film.
A detailed discussion of the effect of long-range intermolecular forces on the \emph{stability} of such a film \citep{Fisher2005,Bandyopadhyay2006} lies beyond the scope of the present work.
%
\section{Conclusions} \label{sec:conclusion}
%
In this work, we investigated the dip-coating flow generated by a plate lifted at constant speed through a stratified bath made of two immiscible liquids, focusing on the case when both of them are entrained on the plate.
We presented first a general formulation within the framework of the lubrication approximation that includes the full (nonlinear) expression for the curvature and the effect of gravity.
Following a similar scheme as \cite{WilsonJEM1982}, we then simplified the problem for small capillary numbers $Ca^{1/3} \ll 1$.
Using this asymptotic formulation, we explored the effect of some of the control parameters on the asymptotic thicknesses of the resulting thin liquid films.
We mostly focused our attention on the viscosity ratio $M$ and the thickness of the floating layer $\Delta H$, as these parameters could be varied more easily in experiments.
For completeness we also showed results for a limited number of values of the surface tension ratio $\Sigma$, although this parameter would be harder to vary experimentally using common liquids.

We could rationalize the numerical results by examining the flow in the narrow zone where the liquid / liquid and liquid / air interfaces get very close, which we dubbed \emph{virtual contact point}.
We showed that the shear stresses at the plate and at the liquid / liquid interface at that point alone are sufficient to predict the thicknesses of the two films, through simple scaling laws derived from the original ideas of \cite{LandauLevich1942}.
Regarding the influence of the different control parameters of the problem, we found that the thickness $\hat{h}_{1,\infty}$ of the lowermost film is barely affected by the viscosity ratio $M$.
This is a consequence of the shear stress at the interface separating the two liquids being negligible as compared to the one at the plate / lower liquid interface.
On the contrary, the thickness of the floating layer, $\Delta H$, has a strong impact on $\hat{h}_{1,\infty}$, which grows monotonically with $\Delta H$. 
In this two-liquid configuration, the thickness of the lower film is always larger than the corresponding thickness for a one-liquid Landau-Levich flow.
The thickness $\delta\hat{h}_\infty$ of the uppermost film exhibits a comparatively more complex behaviour, showing a non-monotonous trend with both $M$ and $\Delta H$, with a maximum for a given pair of these parameters. 
More importantly, there is a finite range of values in the ($M$, $\Delta H$) parameter space where $\delta\hat{h}_\infty$ takes physically-realizable values, which amounts to say, where a solution with two entrained films exists.

In summary, we provided evidence that a dip-coating configuration with two entrained films is feasible using existing liquids and we developed the framework to understand and predict the corresponding entrained thicknesses.
The present theory of plate coating could also be extended to fiber coating provided the curvatures of the static menisci, near the fiber and in perfect wetting conditions, are modified to account for the second principal radius of curvature contributing primarily to the capillary suction mechanism \citep{Quere1999}.
In the limit of a fiber radius $b \ll \ell_c$, the adaptation is straightforward and consists in replacing $\sqrt{2}/ \ell_c$ by $1/b$ in the expressions for the two static menisci.

From the point of view of applications, two improvements for industrial dip-coating processes could be envisioned based on the present findings:
\begin{enumerate}
    \item Since adding a floating liquid layer has been shown to increase the thickness of the lower coated layer, this double-layer configuration could be used to reduce the number of passes in multi-pass dip-coating processes \citep{Li1993,Petropoulos1997}, without having to add undesired additives such as surfactants or nanoparticles.
    \item Since the upper film is up to $10^{3}$ times thinner than lower one, this configuration could be used to deposit thin layers of a very viscous fluid (\textit{e.g.} polymers), which would be impossible otherwise, at least at speeds compatible with industrial production. Naturally, this would require the removal of the lower layer, for example by taking advantage of a porous substrate \citep{Aradian2000}. This could be of interest in the fabrication of enhanced textile \citep{Hu2016} or porous membranes \citep{Jesswein2018}.
\end{enumerate} \bigskip
L.C. acknowledges funding from the European Union’s Horizon 2020 research and innovation programme under the Marie Sk\l odowska-Curie grant agreement No 882429.
B.S. thanks the F.R.S-FNRS for financial support.
A.A.K. thanks Universidad Carlos III de Madrid and Banco Santander for the financial support of his Chair of Excellence. 
J.R.-R. acknowledges the support of the Spanish Ministry of Economy and Competitiveness through grants DPI2017-88201-C3-3-R and DPI2018-102829-REDT, partly funded with European funds.  \bigskip \\
Declaration of Interests. The authors report no conflict of interest.
%
%
\appendix
\section{Velocity field and shear stress} \label{apdx:velocity_and_shear}
%
Here we provide expressions for the velocity field, pressure gradients and shear stresses used to compute the data shown in figures \ref{fig:streamlines_shear_pressure_grad} and \ref{fig:plot_vs_tau}.
The vertical velocities, $\uhat_i$, arise from solving equations \eqref{eq:momentum_adim_1} and \eqref{eq:momentum_adim_2} with boundary conditions \eqref{eq:BCx_adim_1} -- \eqref{eq:BCx_adim_4}, while the horizontal ones, $\vhat_i$, come from solving the continuity equation.
In the following expressions, a hat is used to denote rescaled variables, as introduced in § \ref{ssec:rescaling}.
%
\subsubsection*{Velocity field in fluid 1}
%
\begin{equation} \label{eq:u1}
    \uhat_1 = u_1 = \hPi_1\left(\frac{\xhat^2}{2} -  \hhat_1 \xhat\right) - \hPi_2 \xhat \left(\hhat_2-\hhat_1\right) + 1
\end{equation}
\begin{equation} \label{eq:v1}
    \vhat_1 = v_1\,Ca^{-1/3} = \hPi'_1\left(-\frac{\xhat^3}{6} + \hhat_1\frac{\xhat^2}{2}\right) +  \hPi'_2\frac{\xhat^2}{2}\left(\hhat_2-\hhat_1\right) + \hPi_1\frac{\xhat^2}{2}\hhat_1 + \hPi_2\frac{\xhat^2}{2}\left(\hhat'_2-\hhat'_1\right)
\end{equation}
%
\subsubsection*{Velocity field in fluid 2}
%
\begin{equation} \label{eq:u2}
    \uhat_2 = u_2 = -\hPi_1\frac{\hhat_1^2}{2} + \hPi_2\left[\frac{1}{M}\left(\frac{\xhat^2}{2} - \hhat_2\xhat - \frac{\hhat_1^2}{2} + \hhat_1\hhat_2\right) - \hhat_1\left(\hhat_2-\hhat_1\right)\right]+1
\end{equation}

\begin{multline} \label{eq:v2}
        \vhat_2 = v_2\,Ca^{-1/3} = \hPi'_1 \xhat\frac{\hhat_1^2}{2} - \frac{\hPi'_2}{M}\left[\frac{\xhat^3}{6} - \frac{\hhat_2\xhat^2}{2} + \xhat\left(-\frac{\hhat_1^2}{2} + \hhat_1\hhat_2 - M\hhat_1\left(\hhat_2-\hhat_1\right)\right)
    \right] +\\
    \hPi_1 \xhat \hhat_1 \hhat'_1 -
    \frac{\hPi_2}{M}\left[-\hhat'_2\frac{\xhat^2}{2} + \xhat\left(-\hhat_1\hhat'_1 + \left(\hhat_1\hhat_2\right)' -
     M\left(\hhat_1\left(\hhat_2-\hhat_1\right)\right)'
     \right)\right] + C(\zhat)
\end{multline}
The function $C(\zhat)$ is the one that guarantees $\vhat_1(\zhat, \hhat_1) = \vhat_2(\zhat, \hhat_1)$. 
In these equations, the prime denotes the derivative with respect to $\hat z$.
Moreover the linearized rescaled pressure gradients $\hPi_1$ and $\hPi_2$ and their derivatives with respect to $\zhat$, $\hPi'_1$ and $\hPi'_2$, are given by:
\begin{equation}
    \hPi_1 = \lim_{Ca \rightarrow 0} Ca^{1/3} \Pi_1 = -\hhat_1''' - \Sigma \hhat_2'''
    \label{eq:def_hPi1}
\end{equation}
\begin{equation}
    \hPi_2 = \lim_{Ca \rightarrow 0} Ca^{1/3} \Pi_2 = -\Sigma \hhat_2'''
    \label{eq:def_hPi2}
\end{equation}
\begin{equation}
    \hPi'_1 = -\hhat_1'''' - \Sigma \hhat_2''''
    \label{eq:def_gradPi1}
\end{equation}
\begin{equation}
    \hPi'_2 = -\Sigma \hhat_2''''
    \label{eq:def_gradPi2}
\end{equation}
Note that taking the limit in expressions (\ref{eq:def_hPi1}) and (\ref{eq:def_hPi2}) amounts to simultaneously neglecting the gravity term in the definitions of the pressure gradients $\Pi_{i}$ (equations (\ref{eq:gradP_adim_1}) -- (\ref{eq:gradP_adim_2})) and to linearize the curvature, \textit{i.e.} to set $\kappa_{i} \approx h_{i}''$.\\

\noindent Finally, the shear stresses at the plate / liquid 1 interface, $\htau_{01}$, and at the liquid 1 / liquid 2 interface, $\htau_{12}$, are:
\begin{equation}
    \htau_{01} = Ca^{-1/3} \tau_{xz}(x=0) = -\hPi_1\,\hhat_1 - \hPi_2\left(\hhat_2-\hhat_1\right),
    \label{eq:def_htau0}
\end{equation}
\begin{equation}
    \htau_{12} = Ca^{-1/3} \tau_{xz}(x=h_1) = -\hPi_2\left(\hhat_2-\hhat_1\right).
    \label{eq:def_htau1}
\end{equation}
In figure \ref{fig:streamlines_shear_pressure_grad}b and c, we also show the shear stress at the plate and the derivative of the pressure gradient, corresponding to the one-liquid Landau-Levich-Derjaguin problem.
Their respective expressions are
\begin{equation}
    \htau_{\rm LLD} = \hhat_1 \hhat_1''' \qquad \text{and} \qquad%
    \hPi_{\rm LLD} = - \hhat_1'''.
    \label{eq:def_htauLLD}
\end{equation}
%
\section{Time-dependent formulation and numerical method} \label{apdx:transient_and_numerics}
%
Although we ultimately seek for steady-state solutions of the problem, as described in section \ref{sec:model_description}, we develop here a \emph{quasi-steady} formulation (§ \ref{ssec:transient_formulation}), where the unsteady terms are kept in the mass conservation equations.
This formulation allowed us to find the steady numerical solutions for $h_1$ and $h_2$ by time marching (§ \ref{ssec:numerical_method}).
The fact that we reach the steady solution by time-marching ensures that the steady state is stable.
%
\subsection{Time-dependent formulation} \label{ssec:transient_formulation}
%
In the framework of our quasi-steady formulation, the dimensional thickness-averaged continuity equations \eqref{eq:continuity_dim_1and2} are replaced by
\begin{equation}
    \frac{\partial h_i}{\partial t} + \frac{\partial Q_i}{\partial z} = 0, %
    \quad \quad \mathrm{for} \ i = 1, 2.
    \label{eq:continuity_dim_unsteady}
\end{equation}
while the momentum conservation equations \eqref{eq:momentum_dim_1} -- \eqref{eq:momentum_dim_2} remain unchanged.
Non-dimensionalization is performed as explained in § \ref{ssec:non_dimensionalization}, making the time dimensionless with the capillary time, namely
\begin{equation}
t \quad\longrightarrow\quad t \, \ell_c\mu_1/\sigma_{12}.
\end{equation}
Equations \eqref{eq:continuity_dim_unsteady} are unaltered upon non-dimensionalization and the flow rates $Q_i$ are still given by \eqref{eq:Q12}.
Using the rescaling and simplifications developed in § \ref{ssec:rescaling}, the quasi-steady equations satisfied by the film thicknesses at leading order are 
\begin{equation}
\frac{\partial \hat{h}_i}{\partial t} + \frac{\partial}{\partial \hat{z}}
\left[\hat{h}_i + \frac{\partial^3 \hat{h}_1}{\partial \hat{z}^3}\hat{F}_{i1} +%
\Sigma \, \frac{\partial^3 \hat{h}_2}{\partial \hat{z}^3}\left(\hat{F}_{i1} + \hat{F}_{i2}\right)\right] = 0,\label{eq:thinfilm_eqns_unsteady}
\end{equation}
where the $\hat{F}_{ij}$ are the same as in the steady formulation (equations \eqref{eq:F11} -- \eqref{eq:F22}).
%
\subsection{Numerical method} \label{ssec:numerical_method}
%
The system of equations (\ref{eq:thinfilm_eqns_unsteady}) with the boundary conditions described in § \ref{ssec:rescaling} and § \ref{ssec:matching} for $\hat{z} \rightarrow \pm\infty$ are discretized in space using first-order one-dimensional finite volumes in a staggered grid. The thicknesses are defined at nodes placed at the center of the elements, while the flow rates,
\begin{equation}
    \qhat_i = \hat{h}_i + \frac{\partial^3 \hat{h}_1}{\partial \hat{z}^3}\hat{F}_{i1} +%
\Sigma \, \frac{\partial^3 \hat{h}_2}{\partial \hat{z}^3}\left(\hat{F}_{i1} + \hat{F}_{i2}\right),
\end{equation}
are defined at nodes located at their boundaries. The resulting set of ordinary differential equations for the discretized thicknesses, $\hhat_i$, is then time-marched with the routine {\tt odeint} implemented in the scientific package SciPy of Python.

To impose the upstream boundary conditions as we approach the static meniscus, which in the scaled variables is equivalent to $\hat{z} \rightarrow -\infty$, we set $\hhat_1$ to a large value, say $\hhat_{1,\rmb} = 100$, and then we vary $\hhat_2$ at that boundary, $\hhat_{2,\rmb}$, along a range of values larger than $\hhat_{1,\rmb}$. Notice that, numerically, we can impose these boundary conditions at $\hat z=0$ without loss of generality, thanks to the translational invariance of the problem.
For every pair $(\hhat_{1,\rmb}, \hhat_{2,\rmb})$ we can then compute $\Delta \zhat_{cl} = \zhat_{cl,2} - \zhat_{cl,1}$ using equations (\ref{eq:eta1_parabola}) -- (\ref{eq:eta2_parabola}) and, finally, relate this parameter to $\Delta H$ using equations (\ref{eq:kappa_stat_1}) -- (\ref{eq:kappa_stat_2}). 

Besides imposing $\hhat_1$ and $\hhat_2$ at the upstream boundary of the numerical domain, we also need to impose the second derivatives (taken from the static menisci solution) there, for which purpose we use a non-centered finite-difference scheme. 
The same approach is used to impose the boundary conditions at the downstream boundary (corresponding physically to $\zhat \rightarrow \infty$), where we set $\partial\hhat/\partial\zhat = \partial^2\hhat/\partial\zhat^2 = 0$. In our numerical method, this boundary condition is applied at $\zhat = 60$, a value sufficiently large so that the results do not depend on it.

As for the initial conditions, although the formulation introduced in this paper is able to describe transient phenomena, here we are only interested in the long-term, steady solution. For this reason we start the time-marching procedure from initial conditions that do not represent any actual physical configuration but that satisfy the boundary conditions described above. In particular, we use a linear function of decaying exponentials, what ensures a smooth start-up of the time-marching procedure, namely,
\begin{eqnarray}
\hat{h}_1(\hat{z}, 0) & = & \sqrt{2(1-R)} \exp{(-L\hat{z})}/L^2,\\
\hat{h}_2(\hat{z}, 0) & = & \sqrt{\frac{2R}{\Sigma}} \exp{(-L\hat{z})}/L^2,
\end{eqnarray}
where the constant $L$ has the value $L=1$ for all the simulations reported here.
%
\section{One-liquid Landau-Levich \textit{vade mecum}} \label{apdx:one_liquid_LLD}
%
In this appendix, we derive some properties of the classical one-liquid dip-coating flow \citep{LandauLevich1942} that are relevant for the discussion of its two-liquid counterpart.
For a single-phase LLD flow, the film thickness $\hhat$ obeys the differential equation
\begin{equation}
    \hhat_\infty = \hhat + s \hhat'''\frac{\hhat^3}{3},
\end{equation}
where $s$ represents a dimensionless surface tension.
For instance, if we wish to apply this equation to interface (I), $s = 1$, while for interfaces (II) and (III) (see figure \ref{fig:sketch_contact_point}) it would be $s = \Sigma$ and $s = 1 + \Sigma$, respectively.
To write this equation, the thickness $\hhat$ and streamwise coordinate $\zhat$ have been made dimensionless with a capillary length times $Ca^{2/3}$ and $Ca^{1/3}$, respectively. 
Using this notation, $\hhat_\infty$ represents the dimensionless flat film thickness, far above the bath, and is equivalent also to the dimensionless flow rate transported by the film.

We start by proving that the approximate film curvature, $\hhat''$, decays monotonically with $\zhat$, \textit{i.e.} moving up along the plate. 
In a first step, we introduce the change of variables $\eta = \hhat/\hhat_\infty$ and $\xi = \zhat/(\hhat_\infty s^{1/3})$, which yields
\begin{equation}
    1 = \eta + \eta'''\frac{\eta^3}{3},
    \label{eq:LLD_one_liquid_eta}
\end{equation}
where the prime denotes now the derivative with respect to $\xi$.
In a second step, as suggested in \cite{LandauLevich1942}, we take advantage of the autonomous character of the equation to reduce its order through the substitution $\eta' = -F^{1/2}$.
In terms of this function $F$, the interface curvature becomes $\eta'' = \frac{1}{2}\mathrm{d}F/\mathrm{d}\eta$ and equation (\ref{eq:LLD_one_liquid_eta}) turns into
\begin{equation}
    \frac{\mathrm{d}^2F}{\mathrm{d}\eta^2} = \frac{6\left(\eta-1\right)}{\eta^3 F^{1/2}}.
    \label{eq:LLD_transformed_F}
\end{equation}
Since, by definition, $\eta > 1$, this equation reveals that $\mathrm{d}^2F/\mathrm{d}\eta^2 > 0$ and therefore that $\mathrm{d}F/\mathrm{d}\eta = 2\eta''$ grows monotonically with the film thickness $\eta$. 
Likewise, since the film thickness decreases monotonically with the height $\zhat$, as $\eta' = -F^{1/2} < 0$, we can conclude that the interfacial curvature $\eta''$ also decreases monotonically as we move up along the plate.

Regarding the flat film thickness, $\hhat_\infty$, it is possible to relate it to the curvature in the limit $\eta \gg 1$, that is, where the film connects with the static meniscus.
Integrating equation (\ref{eq:LLD_transformed_F}) with the initial conditions $F = \mathrm{d}F/\mathrm{d}\eta = 0$ at $\eta = 1$, as suggested in \cite{LandauLevich1942}, we get $\mathrm{d}F/\mathrm{d}\eta(\eta \rightarrow \infty) = 2 \eta''(\eta \rightarrow \infty) = 2.673$. 

In terms of the original variables $\hhat$ and $\zhat$,
\begin{equation}
    K = \frac{\rmd^2\hhat}{\rmd\zhat^2}(\zhat \rightarrow -\infty) = \frac{\eta''(\eta \rightarrow \infty)}{s^{2/3} \hhat_\infty},
\end{equation}
where we denoted the curvature of the static meniscus near the wall by $K$.
In perfect wetting conditions, $K = \sqrt{2}$.
Finally,
\begin{equation}
    \hhat_\infty = \frac{1.336}{s^{2/3} K}.
    \label{eq:h_infty_vs_K_LLD}
\end{equation}
This result shows that the higher the curvature far away from the flat film region, the thinner the film thickness. 
This has been used in § \ref{ssec:trend_with_DeltaH} to understand the variation of lower film thickness $\hhat_{1, \infty}$ with $\Delta H$ and to estimate a lower bound for this quantity.
%
\bibliography{lld2layers} 

\begin{thebibliography}{44}
\expandafter\ifx\csname natexlab\endcsname\relax\def\natexlab#1{#1}\fi
\def\au#1{#1} \def\ed#1{#1} \def\yr#1{#1}\def\at#1{#1}\def\jt#1{\textit{#1}}
  \def\bt#1{#1}\def\bvol#1{\textbf{#1}} \def\vol#1{#1} \def\pg#1{#1}
  \def\publ#1{#1}\def\arxiv#1{#1}\def\org#1{#1}\def\st#1{\textit{#1}}

\bibitem[Aradian {\em et~al.\/}(2000)Aradian, Raphael \&
  De~Gennes]{Aradian2000}
{\sc \au{Aradian, A.}, \au{Raphael, E.} \& \au{De~Gennes, P.-G.}} \yr{2000}
  \at{Dewetting on porous media with aspiration}.  \jt{The European Physical
  Journal E}  \bvol{2}~(4),  \pg{367--376}.

\bibitem[Bandyopadhyay \& Sharma(2006)]{Bandyopadhyay2006}
{\sc \au{Bandyopadhyay, D.} \& \au{Sharma, A.}} \yr{2006}  \at{Nonlinear
  instabilities and pathways of rupture in thin liquid bilayers}.  \jt{The
  Journal of Chemical Physics}  \bvol{125}~(5),  \pg{054711}.

\bibitem[Brochard-Wyart {\em et~al.\/}(1993)Brochard-Wyart, Martin \&
  Redon]{Brochard1993}
{\sc \au{Brochard-Wyart, F.}, \au{Martin, P.} \& \au{Redon, C.}} \yr{1993}
  \at{Liquid/liquid dewetting}.  \jt{Langmuir}  \bvol{9}~(12),
  \pg{3682--3690}.

\bibitem[Champougny {\em et~al.\/}(2017)Champougny, Rio, Restagno \&
  Scheid]{Champougny2017}
{\sc \au{Champougny, L.}, \au{Rio, E.}, \au{Restagno, F.} \& \au{Scheid, B.}}
  \yr{2017}  \at{The break-up of free films pulled out of a pure liquid bath}.
  \jt{Journal of Fluid Mechanics}  \bvol{811},  \pg{499–524}.

\bibitem[Champougny {\em et~al.\/}(2015)Champougny, Scheid, Restagno, Vermant
  \& Rio]{Champougny2015}
{\sc \au{Champougny, L.}, \au{Scheid, B.}, \au{Restagno, F.}, \au{Vermant, J.}
  \& \au{Rio, E.}} \yr{2015}  \at{Surfactant-induced rigidity of interfaces: a
  unified approach to free and dip-coated films}.  \jt{Soft Matter}  \bvol{11},
   \pg{2758--2770}.

\bibitem[Daniel {\em et~al.\/}(2017)Daniel, Timonen, Li, Velling \&
  Aizenberg]{Daniel2017}
{\sc \au{Daniel, D.}, \au{Timonen, J. V.~I.}, \au{Li, R.}, \au{Velling, S.~J.}
  \& \au{Aizenberg, J.}} \yr{2017}  \at{Oleoplaning droplets on lubricated
  surfaces}.  \jt{Nature Physics}  \bvol{13}~(10),  \pg{1020--1025}.

\bibitem[De~Gennes {\em et~al.\/}(2013)De~Gennes, Brochard-Wyart \&
  Qu{\'e}r{\'e}]{DeGennes2013}
{\sc \au{De~Gennes, P.-G.}, \au{Brochard-Wyart, F.} \& \au{Qu{\'e}r{\'e}, D.}}
  \yr{2013} {\em Capillarity and wetting phenomena: drops, bubbles, pearls,
  waves\/}.  \publ{Springer Science \& Business Media}.

\bibitem[Derjaguin(1943)]{Derjaguin1943}
{\sc \au{Derjaguin, B.~V.}} \yr{1943}  \at{On the thickness of the liquid film
  adhering to the walls of a vessel after emptying}.  \jt{{A}cta {P}hysicochem.
  {U}RSS}  \bvol{20},  \pg{349--352}.

\bibitem[Derjaguin \& Churaev(1978)]{Derjaguin1978}
{\sc \au{Derjaguin, B.~V.} \& \au{Churaev, N.~V.}} \yr{1978}  \at{On the
  question of determining the concept of disjoining pressure and its role in
  the equilibrium and flow of thin films}.  \jt{Journal of Colloid and
  Interface Science}  \bvol{66}~(3),  \pg{389--398}.

\bibitem[Dixit \& Homsy(2013)]{Dixit2013}
{\sc \au{Dixit, H.~N.} \& \au{Homsy, G.~M.}} \yr{2013}  \at{The elastic
  {Landau-Levich} problem}.  \jt{Journal of Fluid Mechanics}  \bvol{732},
  \pg{5}.

\bibitem[Feng {\em et~al.\/}(2016)Feng, Muradoglu, Kim, Ault \&
  Stone]{Feng2016}
{\sc \au{Feng, J.}, \au{Muradoglu, M.}, \au{Kim, H.}, \au{Ault, J.~T.} \&
  \au{Stone, H.~A.}} \yr{2016}  \at{Dynamics of a bubble bouncing at a
  liquid/liquid/gas interface}.  \jt{Journal of Fluid Mechanics}  \bvol{807},
  \pg{324--352}.

\bibitem[Feng {\em et~al.\/}(2014)Feng, Roch{\'e}, Vigolo, Arnaudov, Stoyanov,
  Gurkov, Tsutsumanova \& Stone]{Feng2014}
{\sc \au{Feng, J.}, \au{Roch{\'e}, M.}, \au{Vigolo, D.}, \au{Arnaudov, L.~N.},
  \au{Stoyanov, S.~D.}, \au{Gurkov, T.~D.}, \au{Tsutsumanova, G.~G.} \&
  \au{Stone, H.~A.}} \yr{2014}  \at{Nanoemulsions obtained via bubble-bursting
  at a compound interface}.  \jt{Nature Physics}  \bvol{10}~(8),
  \pg{606--612}.

\bibitem[Fingas(2015)]{Fingas2015}
{\sc \au{Fingas, M.~F.}} \yr{2015} {\em Handbook of oil spill science and
  technology\/}.  \publ{Wiley Online Library}.

\bibitem[Fisher \& Golovin(2005)]{Fisher2005}
{\sc \au{Fisher, L.~S.} \& \au{Golovin, A.~A.}} \yr{2005}  \at{Nonlinear
  stability analysis of a two-layer thin liquid film: Dewetting and autophobic
  behavior}.  \jt{Journal of Colloid and Interface Science}  \bvol{291}~(2),
  \pg{515--528}.

\bibitem[Hardy(1982)]{Hardy1982}
{\sc \au{Hardy, J.~T.}} \yr{1982}  \at{The sea surface microlayer: biology,
  chemistry and anthropogenic enrichment}.  \jt{Progress in Oceanography}
  \bvol{11}~(4),  \pg{307--328}.

\bibitem[Hu(2016)]{Hu2016}
{\sc \au{Hu, J.}} \yr{2016} {\em Active coatings for smart textiles\/}.
  \publ{Woodhead Publishing}.

\bibitem[Hurez \& Tanguy(1990)]{Hurez1990}
{\sc \au{Hurez, P.} \& \au{Tanguy, P.~A.}} \yr{1990}  \at{Finite element
  analysis of dip coating with {Bingham} fluids}.  \jt{Polymer Engineering \&
  Science}  \bvol{30}~(18),  \pg{1125--1132}.

\bibitem[Israelachvili(2011)]{Israelachvili2011}
{\sc \au{Israelachvili, J.~N.}} \yr{2011} {\em Intermolecular and surface
  forces\/}.  \publ{Academic press}.

\bibitem[Jesswein {\em et~al.\/}(2018)Jesswein, Uebele, Dieterich, Keller,
  Hirth \& Schiestel]{Jesswein2018}
{\sc \au{Jesswein, I.}, \au{Uebele, S.}, \au{Dieterich, A.}, \au{Keller, S.},
  \au{Hirth, T.} \& \au{Schiestel, T.}} \yr{2018}  \at{Influence of surface
  properties on the dip coating behavior of hollow fiber membranes}.
  \jt{Journal of Applied Polymer Science}  \bvol{135}~(16),  \pg{46163}.

\bibitem[Jinkins {\em et~al.\/}(2017)Jinkins, Chan, Brady, Gronski, Gopalan,
  Evensen, Berson \& Arnold]{Jinkins2017}
{\sc \au{Jinkins, K.~R.}, \au{Chan, J.}, \au{Brady, G.~J.}, \au{Gronski,
  K.~K.}, \au{Gopalan, P.}, \au{Evensen, H.~T.}, \au{Berson, A.} \& \au{Arnold,
  M.~S.}} \yr{2017}  \at{Nanotube alignment mechanism in floating evaporative
  self-assembly}.  \jt{Langmuir}  \bvol{33}~(46),  \pg{13407--13414}.

\bibitem[Kreder {\em et~al.\/}(2018)Kreder, Daniel, Tetreault, Cao, Lemaire,
  Timonen \& Aizenberg]{Kreder2018}
{\sc \au{Kreder, M.~J.}, \au{Daniel, D.}, \au{Tetreault, A.}, \au{Cao, Z.},
  \au{Lemaire, B.}, \au{Timonen, J. V.~I.} \& \au{Aizenberg, J.}} \yr{2018}
  \at{Film dynamics and lubricant depletion by droplets moving on lubricated
  surfaces}.  \jt{Physical Review X}  \bvol{8}~(3),  \pg{031053}.

\bibitem[Landau \& Levich(1942)]{LandauLevich1942}
{\sc \au{Landau, L.~D.} \& \au{Levich, B.}} \yr{1942}  \at{Dragging of a liquid
  by a moving plate}.  \jt{{A}cta {P}hysicochem. {U}RSS}  \bvol{17},
  \pg{141--153}.

\bibitem[Landau \& Lifshitz(1987)]{LandauLifshitzBOOK}
{\sc \au{Landau, L.~D.} \& \au{Lifshitz, E.~M.}} \yr{1987} {\em Fluid
  Mechanics\/}, 2nd edn.  \publ{New York: Pergamon Press}.

\bibitem[L\'eger \& Joanny(1992)]{Leger1992}
{\sc \au{L\'eger, L.} \& \au{Joanny, J.~F.}} \yr{1992}  \at{Liquid spreading}.
  \jt{Reports on Progress in Physics}  \bvol{55}~(4),  \pg{431}.

\bibitem[Li {\em et~al.\/}(2019)Li, Ueda, Paulssen \& Levkin]{Li2019}
{\sc \au{Li, J.}, \au{Ueda, E.}, \au{Paulssen, D.} \& \au{Levkin, P.~A.}}
  \yr{2019}  \at{Slippery lubricant-infused surfaces: Properties and emerging
  applications}.  \jt{Advanced Functional Materials}  \bvol{29}~(4),
  \pg{1802317}.

\bibitem[Li {\em et~al.\/}(1993)Li, Haertling \& Howng]{Li1993}
{\sc \au{Li, K.~K.}, \au{Haertling, G.~H.} \& \au{Howng, W.-Y.}} \yr{1993}
  \at{An automatic dip coating process for dielectric thin and thick films}.
  \jt{Integrated Ferroelectrics}  \bvol{3}~(1),  \pg{81--91}.

\bibitem[Liss \& Duce(2005)]{Liss2005}
{\sc \au{Liss, P.~S.} \& \au{Duce, R.~A.}} \yr{2005} {\em The sea surface and
  global change\/}.  \publ{Cambridge University Press}.

\bibitem[Maillard {\em et~al.\/}(2016)Maillard, Bleyer, Andrieux, Boujlel \&
  Coussot]{Maillard2016}
{\sc \au{Maillard, M.}, \au{Bleyer, J.}, \au{Andrieux, A.~L.}, \au{Boujlel, J.}
  \& \au{Coussot, P.}} \yr{2016}  \at{Dip-coating of yield stress fluids}.
  \jt{Physics of Fluids}  \bvol{28}~(5),  \pg{053102}.

\bibitem[Ouriemi \& Homsy(2013)]{Ouriemi2013}
{\sc \au{Ouriemi, M.} \& \au{Homsy, G.~M.}} \yr{2013}  \at{Experimental study
  of the effect of surface-absorbed hydrophobic particles on the
  {Landau-Levich} law}.  \jt{Physics of Fluids}  \bvol{25}~(8),  \pg{082111}.

\bibitem[Park(1991)]{Park1991}
{\sc \au{Park, C.-W.}} \yr{1991}  \at{Effects of insoluble surfactants on dip
  coating}.  \jt{Adv. Colloid Interface Sci.}  \bvol{146},  \pg{382--394}.

\bibitem[Petropoulos {\em et~al.\/}(1997)Petropoulos, Foley, Hedrick \&
  Nealey]{Petropoulos1997}
{\sc \au{Petropoulos, M.~C.}, \au{Foley, G. M.~T.}, \au{Hedrick, R.~W.} \&
  \au{Nealey, R.~H.}} \yr{1997} Multiple dip coating method. US Patent
  5,633,046.

\bibitem[Qu\'er\'e(1999)]{Quere1999}
{\sc \au{Qu\'er\'e, D.}} \yr{1999}  \at{Fluid coating on a fiber}.  \jt{Annual
  Review of Fluid Mechanics}  \bvol{31}~(1),  \pg{347--384}.

\bibitem[Rio \& Boulogne(2017)]{RioBoulogne2017}
{\sc \au{Rio, E.} \& \au{Boulogne, F.}} \yr{2017}  \at{Withdrawing a solid from
  a bath: How much liquid is coated?}  \jt{Adv. Colloid Interface Sci.}
  \bvol{247},  \pg{100--114}.

\bibitem[de~Ryck \& Qu{\'e}r{\'e}(1998)]{DeRyck1998}
{\sc \au{de~Ryck, Alain} \& \au{Qu{\'e}r{\'e}, David}} \yr{1998}  \at{Gravity
  and inertia effects in plate coating}.  \jt{Journal of Colloid and Interface
  Science}  \bvol{203}~(2),  \pg{278--285}.

\bibitem[Scheid {\em et~al.\/}(2010)Scheid, Delacotte, Dollet, Rio, Restagno,
  Van~Nierop, Cantat, Langevin \& Stone]{Scheid2010}
{\sc \au{Scheid, B.}, \au{Delacotte, J.}, \au{Dollet, B.}, \au{Rio, E.},
  \au{Restagno, F.}, \au{Van~Nierop, E.~A.}, \au{Cantat, I.}, \au{Langevin, D.}
  \& \au{Stone, H.~A.}} \yr{2010}  \at{The role of surface rheology in liquid
  film formation}.  \jt{EPL (Europhysics Letters)}  \bvol{90}~(2),  \pg{24002}.

\bibitem[Scriven(1988)]{Scriven1988}
{\sc \au{Scriven, L.~E.}} \yr{1988}  \at{Physics and applications of dip
  coating and spin coating}.  \jt{MRS Online Proceedings Library Archive}
  \bvol{121}.

\bibitem[Seiwert {\em et~al.\/}(2011)Seiwert, Clanet \& Qu\'ere]{Seiwert2011}
{\sc \au{Seiwert, J.}, \au{Clanet, C.} \& \au{Qu\'ere, D.}} \yr{2011}
  \at{Coating of a textured solid}.  \jt{Journal of Fluid Mechanics}
  \bvol{669},  \pg{55--63}.

\bibitem[Snoeijer {\em et~al.\/}(2008)Snoeijer, Ziegler, Andreotti, Fermigier
  \& Eggers]{Snoeijer2008}
{\sc \au{Snoeijer, J.~H.}, \au{Ziegler, J.}, \au{Andreotti, B.}, \au{Fermigier,
  M.} \& \au{Eggers, J.}} \yr{2008}  \at{Thick films of viscous fluid coating a
  plate withdrawn from a liquid reservoir}.  \jt{Physical review letters}
  \bvol{100}~(24),  \pg{244502}.

\bibitem[Stewart {\em et~al.\/}(2015)Stewart, Feng, Kimpton, Griffiths \&
  Stone]{Stewart2015}
{\sc \au{Stewart, P.~S.}, \au{Feng, J.}, \au{Kimpton, L.~S.}, \au{Griffiths,
  I.~M.} \& \au{Stone, H.~A.}} \yr{2015}  \at{Stability of a bi-layer free
  film: simultaneous or individual rupture events?}  \jt{Journal of Fluid
  Mechanics}  \bvol{777},  \pg{27--49}.

\bibitem[Takamura {\em et~al.\/}(2012)Takamura, Fischer \&
  Morrow]{Takamura2012gly}
{\sc \au{Takamura, K.}, \au{Fischer, H.} \& \au{Morrow, N.~R.}} \yr{2012}
  \at{Physical properties of aqueous glycerol solutions}.  \jt{Journal of
  Petroleum Science and Engineering}  \bvol{98},  \pg{50--60}.

\bibitem[Tewes {\em et~al.\/}(2019)Tewes, Wilczek, Gurevich \&
  Thiele]{Tewes2019}
{\sc \au{Tewes, W.}, \au{Wilczek, M.}, \au{Gurevich, S.~V.} \& \au{Thiele, U.}}
  \yr{2019}  \at{Self-organized dip-coating patterns of simple, partially
  wetting, nonvolatile liquids}.  \jt{Physical Review Fluids}  \bvol{4}~(12),
  \pg{123903}.

\bibitem[Vella {\em et~al.\/}(2004)Vella, Aussillous \& Mahadevan]{Vella2004}
{\sc \au{Vella, D.}, \au{Aussillous, P.} \& \au{Mahadevan, L.}} \yr{2004}
  \at{Elasticity of an interfacial particle raft}.  \jt{EPL (Europhysics
  Letters)}  \bvol{68}~(2),  \pg{212}.

\bibitem[Vollhardt \& Fainerman(2010)]{Vollhardt2010}
{\sc \au{Vollhardt, D.} \& \au{Fainerman, V.~B.}} \yr{2010}
  \at{Characterisation of phase transition in adsorbed monolayers at the
  air/water interface}.  \jt{Advances in Colloid and Interface Science}
  \bvol{154}~(1-2),  \pg{1--19}.

\bibitem[Wilson(1982)]{WilsonJEM1982}
{\sc \au{Wilson, S. D.~R.}} \yr{1982}  \at{The drag-out problem in film coating
  theory}.  \jt{J. Eng. Math.}  \bvol{16},  \pg{209--221}.

\end{thebibliography}
\bibliographystyle{jfm} 
\end{document}